\documentclass[aps,superscriptaddress,prc,twocolumn,nofootinbib,showpacs,showkeys]{revtex4-1}
\usepackage{graphicx}
\usepackage{epstopdf}
\usepackage{subfigure}
\usepackage{epsfig}
\usepackage{amsmath,amssymb,amsfonts}
\usepackage[utf8]{inputenc}
\usepackage{color}
\usepackage[bookmarks,
                bookmarksopen = true,
                bookmarksnumbered = true,
                linktocpage,
                colorlinks = true,
                linkcolor = blue,
                urlcolor  = blue,
                citecolor = blue,
                anchorcolor = green,
                hyperindex = true,
                hyperfigures]
                {hyperref}

\begin{document}

\title{Heavy flavor quenching and flow: the roles of initial condition, pre-equilibrium evolution and in-medium interaction}

\author{Shu-Qing Li}
\email{lisq79@jnxy.edu.cn}
\affiliation{Department of Physics and Information Engineering, Jining University, Qufu, Shandong, 273155, China}
\affiliation{Institute of Particle Physics and Key Laboratory of Quark and Lepton Physics (MOE), Central China Normal University, Wuhan, Hubei, 430079, China}
\author{Wen-Jing Xing}
\affiliation{Institute of Particle Physics and Key Laboratory of Quark and Lepton Physics (MOE), Central China Normal University, Wuhan, Hubei, 430079, China}
\author{Feng-Lei Liu}
\affiliation{Institute of Particle Physics and Key Laboratory of Quark and Lepton Physics (MOE), Central China Normal University, Wuhan, Hubei, 430079, China}
\author{Shanshan Cao}
\email{sshan.cao@gmail.com}
\affiliation{Cyclotron Institute, Texas A\&M University, College Station, Texas, 77843, USA}
\affiliation{Department of Physics and Astronomy, Wayne State University, 666 W. Hancock St., Detroit, Michigan 48201, USA}
\author{Guang-You Qin}
\email{guangyou.qin@mail.ccnu.edu.cn}
\affiliation{Institute of Particle Physics and Key Laboratory of Quark and Lepton Physics (MOE), Central China Normal University, Wuhan, Hubei, 430079, China}
\affiliation{Nuclear Science Division, Lawrence Berkeley National Laboratory, Berkeley, CA 94720, USA}
\date{\today}


\begin{abstract}

Within an advanced Langevin-hydrodynamics framework coupled to a hybrid fragmentation-coalescence hadronization model, we study heavy flavor quenching and flow in relativistic heavy-ion collisions. We investigate how the initial heavy quark spectrum, the energy loss and hadronization mechanisms of heavy quarks in medium, the evolution profile of pre-equilibrium stage, the flow of medium and the temperature dependence of heavy quark diffusion coefficient influence the suppression and elliptic flow of heavy mesons at RHIC and the LHC.
Our result shows that different modeling of initial conditions, pre-equilibrium evolution and in-medium interaction can individually yield about 10-40\% uncertainties in $D$ meson suppression and flow at low transverse momentum.
We also find that a proper combination of collisional versus radiative energy loss, coalescence versus fragmentation in hadronization, and the inclusion of medium flow are the most important factors for describing the suppression and elliptic flow of heavy mesons.

\end{abstract}

\maketitle


\section{Introduction}
\label{sec:Introduction}

Relativistic heavy-ion collisions provide a unique opportunity to study nuclear matter under extreme density and temperature.
It is now generally acknowledged that a color-deconfined QCD matter, known as Quark-Gluon Plasma (QGP), has been produced in high-energy nuclear collisions at the Relativistic Heavy-Ion Collider (RHIC) and the Large Hadron Collider (LHC)~\cite{Shuryak:2014zxa}.
The QGP behaves like a strongly-interacting fluid, as revealed by the large anisotropic collective flow of hadrons emitted in these energetic heavy-ion collisions~\cite{Adams:2003zg,Aamodt:2010pa,ATLAS:2011ah}.
The collective flow has been successfully explained by relativistic hydrodynamic simulation with small values of shear-viscosity-to-entropy-density ratio ($\eta/s$)~\cite{Gale:2013da, Heinz:2013th,Song:2010mg,Bernhard:2019bmu}.

Another evidence of quark-gluon degrees of freedom inside the QGP is jet quenching~\cite{Wang:1991xy,Qin:2015srf,Cao:2020wlm}.
Energetic quarks and gluons produced in the primordial stage of nuclear collisions lose energy while traversing the hot and dense nuclear matter before fragmenting into hadrons.
The study of energetic jets and hadrons and their medium modification in heavy-ion collisions provides a valuable probe of the QGP properties, such as the jet transport coefficient inside the QGP~\cite{Burke:2013yra}.
A lot of effort has been devoted to understanding the nuclear modification of jets, such as single inclusive hadron/jet suppression~\cite{Burke:2013yra,Chang:2016gjp,Cao:2017qpx,He:2018xjv,Casalderrey-Solana:2018wrw}, di-hadron/jet correlations~\cite{Zhang:2007ja,Qin:2010mn,Cao:2015cba,Majumder:2004pt,Chen:2016cof}, photon/$Z$-triggered hadron/jet correlations~\cite{Qin:2009bk,Zhang:2009rn,Qin:2012gp,Chen:2016vem,Chen:2017zte,Luo:2018pto}, and more differential substructures of full jets~\cite{Casalderrey-Solana:2016jvj,Tachibana:2017syd,Park:2018acg,Chang:2019sae,Chang:2017gkt}.
Among various hard probes, heavy quarks are of particular interest since their flavors are conserved when interacting with QGP. Thus, they can serve as a clean probe of the traversed nuclear medium~\cite{Dong:2019byy,Dong:2019unq}.
At low $p_\mathrm{T}$, heavy quarks may probe the color potential of the QGP~\cite{Liu:2018syc}; at intermediate $p_\mathrm{T}$, heavy flavor hadron chemistry can help constrain our knowledge about the hadronization process of jet partons~\cite{Song:2018tpv,Plumari:2017ntm,He:2019vgs,Cho:2019lxb,Cao:2019iqs}; at high $p_\mathrm{T}$, heavy quarks can directly probe the mass hierarchy of parton energy loss inside a thermalized QGP medium~\cite{Cao:2017hhk,Xing:2019xae}.

Various transport models have been developed to study heavy quark dynamics in heavy-ion collisions. Some models only include collisional energy loss of heavy quarks assuming the large mass approximation~\cite{He:2011qa,Das:2015ana,Song:2015ykw}; others also consider radiative energy loss and medium-induced gluon emission which are important for high $p_\mathrm{T}$ heavy quarks~\cite{Gossiaux:2006yu,Gossiaux:2010yx,Das:2010tj,Fochler:2013epa,Cao:2013ita,Cao:2015hia,Cao:2016gvr,Cao:2017crw,Ke:2018tsh,Li:2019wri,Prado:2019ste}. While various models have been built on rather different assumptions about heavy-quark-medium interaction, many of them can provide reasonable descriptions of experimental data on heavy quarks.
In Ref.~\cite{Cao:2018ews}, it is shown that a factor of 2 difference still remains in the heavy quark transport coefficient when different transport models are tuned to describe the same data.
Note that such discrepancy mainly originates from different energy loss mechanisms, and the uncertainties could be much larger if the effects from using different medium profiles and hadronization models are included.
To achieve a clear picture of heavy flavor production and medium modification, it is important to understand the systematic uncertainties from various model components, such as the initial heavy quark spectrum, the assumptions of heavy-quark-medium interaction in the pre-equilibrium stage, heavy quark energy loss and hadronization mechanisms, the geometry and flow of the QGP, as well as the temperature dependence of heavy quark transport coefficient. Some of these aspects have been discussed in earlier studies~\cite{Rapp:2018qla,Cao:2018ews,Xu:2018gux,Andres:2019eus,Katz:2019fkc}. In this work, we extend such studies and perform a systematic investigation on the uncertainties from all the above mentioned sources using the state-of-the-art Langevin-hydrodynamics model, which incorporates both collisional and radiative energy loss of heavy quarks through a dynamical QGP medium~\cite{Cao:2013ita,Cao:2015hia}. By combining with the up-to-date fragmentation-coalescence hadronization approach~\cite{Cao:2019iqs}, we analyze the nuclear modification factor ($R_\mathrm{AA}$) and elliptic flow coefficient ($v_2$) of heavy mesons and compare them to experimental data at RHIC and the LHC.

The paper is organized as follows. In Sec.~\ref{sec:model}, we review the production, energy loss and hadronization of heavy quarks within our advanced Langevin-hydrodynamics framework coupled to a hybrid fragmentation-coalescence hadronization model. In Sec.~\ref{sec:results}, we present our numerical results on $D$ meson suppression and elliptic flow, together with their dependences on the initial heavy quark spectrum, the energy loss and hadronization mechanisms, the pre-equilibrium temperature profile of the medium, the flow of medium and the temperature dependence of heavy quark diffusion coefficient. A summary is given in Sec.~\ref{sec:summary}.

\section{Production, energy loss and hadronization of heavy quarks}
\label{sec:model}

\subsection{Initial production of heavy quarks}
\label{subsec:initial}

Because of their large masses, heavy quarks are mainly produced via hard scatterings in the primordial stage of relativistic heavy-ion collisions. This allows us to use a Monte-Carlo (MC) Glauber model to determine the spatial distribution for the production vertices of heavy quarks, and use the perturbative QCD (pQCD) approach to calculate their initial momentum spectra.
In this work, we use the Fixed-Order-Next-to-Leading-Log (FONLL) framework~\cite{Cacciari:2001td,Cacciari:2012ny,Cacciari:2015fta} to calculate the initial heavy quark spectra, if not otherwise specified. Within this framework, we apply the CT14NLO~\cite{Dulat:2015mca} parton distribution function (PDF) for the free proton and the EPPS16~\cite{Eskola:2016oht} parametrization for PDF in nuclei to take into account the nuclear shadowing effect. We will investigate the systematic uncertainties between using these FONLL spectra and the Leading-Order (LO) pQCD spectra~\cite{Combridge:1978kx} in the next section. The nuclear shadowing effect on heavy flavor observables will also be discussed.

\subsection{In-medium evolution of heavy quarks}

During their propagation through the QGP fireball, heavy quarks lose energy via both quasi-elastic scatterings with thermal light partons in medium and the inelastic medium-induced gluon emission~\cite{Wang:1991xy,Braaten:1991we}. In this work, we utilize the following modified Langevin equation~\cite{Cao:2013ita} which simultaneously incorporates these two processes to describe the time evolution of the energy and momentum of heavy quarks while they traverse QGP:
 \begin{equation}
 \label{eq:Langevin}
   \frac{d\vec{p}}{dt} = -\eta _{D}(p)\vec{p}+\vec{\xi}+\vec{f_{g}}.
 \end{equation}
In the above equation, the first two terms on the right-hand side are inherited from the classical Langevin equation, representing the drag force and thermal random force experienced by a heavy quark while it diffuses inside a thermal medium due to multiple scatterings. For a minimal model, we assume the thermal force $\vec{\xi}$ does not depend on the heavy quark momentum and satisfies the correlation relation of a white noise $\langle\xi^{i}(t)\xi^{j}(t^{\prime})\rangle=\kappa\delta^{ij}\delta(t-t^{\prime})$, where $\kappa$ is the momentum diffusion coefficient of heavy quarks and related to the spatial diffusion coefficient via $D_\mathrm{s}\equiv T/[M\eta_{D}(0)]=2T^{2}/\kappa$ if the fluctuation-dissipation relation $\eta_{D}(p)=\kappa/(2TE)$ is respected.

Apart from the above two terms from quasi-elastic scatterings, a third term $\vec{f_{g}}=-d\vec{p}_{g}/dt$ is introduced to describe the recoil force exerted on heavy quarks while they emit medium-induced gluons, with $\vec{p}_{g}$ being the momentum of the emitted gluons. The probability of gluon radiation during the time interval $(t,t+\Delta t)$ is related to the average number of radiated gluons in $\Delta t$ as:
 \begin{equation}
 \label{eq:gluonProb}
 P_\mathrm{rad}(t,\Delta t) = \langle N_{g}(t,\Delta t)\rangle = \Delta t\int dxdk_{\perp}^{2}\frac{dN_{g}}{dxdk_{\perp}^{2}dt}.
 \end{equation}
As long as $\Delta t$ is chosen sufficiently small, the average number $\langle N_{g}(t,\Delta t)\rangle$ is less than 1 and can be interpreted as a probability. In this study, the gluon distribution function in Eq.~(\ref{eq:gluonProb}) is taken from the higher-twist energy loss calculation~\cite{Guo:2000nz,Majumder:2009ge,Zhang:2003wk}:
\begin{equation}
\label{eq:gluonSpectrum}
\frac{dN_{g}}{dxdk_{\perp}^{2}dt}=\frac{2\alpha_{s}P(x)\hat q}{\pi k_{\perp}^{4}}\sin^{2}\left(\frac{t-t_{i}}{2\tau _{f}}\right)\left(\frac{k_{\perp}^{2}}{k_{\perp}^{2}+x^{2}M^{2}}\right)^{4},
\end{equation}
where $x$ is the fractional energy of the emitted gluon taken from the parent heavy quark, $k_\perp$ is the transverse momentum of the gluon, $\alpha _\mathrm{s}$ is the strong coupling which runs with $k_\perp^2$, $P(x)$ is the $Q\rightarrow Qg$ splitting function, and $\tau _{f}=2Ex(1-x)/(k_{\perp}^{2}+x^{2}M^{2})$ is the gluon formation time with $E$ and $M$ being the energy and mass of heavy quarks. Note that the multiplicative term at the end of Eq.~(\ref{eq:gluonSpectrum}) is known as the ``dead cone factor", characterizing the mass dependence of the radiative energy loss of hard partons.
In Eq.~(\ref{eq:gluonSpectrum}), $\hat q$ is the gluon transport coefficient and may be related to the quark diffusion coefficient via $\hat q = 2\kappa C_{A}/C_{F}$. Thus in this modified Langevin model, there is only one free parameter which we choose to be the dimensionless quantity $D_\mathrm{s}(2\pi T)$.

When simulating the radiative energy loss of heavy quarks, a lower cut-off energy of the radiated gluon $\omega _{0}=\pi T$ is imposed to mimic the balance between gluon emission and absorption processes around the thermal scale. Below $\omega _{0}$, the gluon radiation is disabled and the evolution of heavy quarks at low energy is entirely controlled by quasi-elastic scatterings. In other words, $x\in [\pi T/E,1]$ is used when calculating the gluon radiation probability in Eq.~(\ref{eq:gluonProb}). This allows an approximate thermal equilibration of heavy quarks after sufficiently long evolution time although the exact fluctuation-dissipation relation cannot be guaranteed due to the lack of the gluon absorption process~\cite{Cao:2013ita}.

To calculate heavy quark energy loss inside a realistic QGP medium, we couple this improved Langevin approach to a hydrodynamic model that provides the temperature profiles of the QGP fireball. In this work, we use two hydrodynamic models -- (2+1)-dimensional Vishnew~\cite{Song:2007fn,Qiu:2011hf,Song:2007ux} and (3+1)-dimensional CLVisc~\cite{Pang:2012he,Pang:2014ipa} to calculate heavy quark observables. The Glauber model is used to calculate the initial entropy/energy density distributions for these hydrodynamic simulations. The starting time of the hydrodynamic evolution $\tau_{\rm hydro}=0.6$~fm and the shear-viscosity-to-entropy-density-ratio $\eta/s=0.08$ are fixed in order to provide satisfactory soft hadron spectra observed at RHIC and the LHC. At every time step, each heavy quark is first boosted into the local rest frame of the fluid cell through which it propagates. In this frame, the energy and momentum of the heavy quark are updated based on our improved Langevin equation. Then the heavy quark is boosted back to the global center of momentum frame, in which it streams to the next time step. Unless otherwise specified, the starting time $\tau_0$ for heavy-quark-medium interaction is set at $\tau_0=0.6$~fm, which means that we assume free-streaming for heavy quarks before $\tau_0$. However, the uncertainties of heavy flavor observables due to different choices of  $\tau_0$ and different assumptions of the medium temperature profiles before $\tau_0$ will be explored later in this work.

\subsection{Hadronization of heavy quarks}

When the local temperature of hydrodynamic medium drops to $T_\mathrm{c}=160$~MeV, we switch off the interaction between heavy quarks and medium. Then heavy quarks are converted into heavy flavor hadrons by applying our up-to-date hybrid fragmentation-coalescence model~\cite{Cao:2019iqs}.

Typically, fragmentation dominates at high transverse momentum and the heavy-light-quark-coalescence dominates at low transverse momentum.
The momentum-dependent coalescence probability is determined by the wavefunction overlap between the free constituent quark states and the hadronic bound states, and can be expressed as the Wigner function~\cite{Oh:2009zj}. For the coalescence of two quarks into a meson, the Wigner function reads:
\begin{equation}
\label{eq:Wigner}
f^W_M(\vec{r},\vec{q})=g_M\int d^3r' e^{-i\vec{q}\cdot\vec{r}'}\phi_M(\vec{r}+\frac{\vec{r}'}{2})\phi^*_M(\vec{r}-\frac{\vec{r}'}{2}),
\end{equation}
where $\vec{r}$ and $\vec{q}$ are the relative position and momentum between the two quarks in their center of momentum frame. The statistics factor $g_M$ represents the ratio of the spin-color degrees of freedom between the final-state meson and the initial-state quarks. For instances, it is 1/36 for $D^0$ ground state, but 1/12 for $D^{*0}$. In this work, the simple harmonic oscillator potential is assumed for the meson wavefunction $\phi_M$, with the oscillator frequency $
\omega$ as the single parameter of this coalescence model. For three-quark coalescence into baryon, we first combine two quarks and then combine their center-of-momentum with the third quark. Following Ref.~\cite{Cao:2019iqs}, we include both $s$ and $p$-wave hadronic states, which cover all major charmed hadron species reported by the Particle Data Group (PDG)~\cite{Tanabashi:2018oca}. This is essential in understanding the observed charmed hadron chemistry -- $\Lambda_c/D^0$ and $D_s/D^0$ ratios -- at RHIC and the LHC~\cite{Cao:2019iqs}. 

The above Wigner functions are used to calculate the probabilities for heavy quarks to coalesce with thermal light quarks into all possible hadrons when they cross the QGP boundary. Light quarks are assumed to follow the thermal distribution in the local rest frame of the expanding QGP. Based on these probabilities, we use the Monte-Carlo method to determine which specific coalescence channels heavy quarks take to become hadrons. For heavy quarks that do not coalesce, we implement PYTHIA simulation~\cite{Sjostrand:2006za} to fragment them into hadrons. The above mentioned model parameter $\omega$ is determined such that the total coalescence probability of a zero momentum charm quark is 1 since it is kinematically forbidden to fragment. This yields $\omega=0.24$~GeV after including all $s$ and $p$-wave charmed hadron states.

\section{Nuclear modification of heavy flavor mesons}
\label{sec:results}

In this section, we present our numerical results on the nuclear modifications of heavy flavor mesons in heavy-ion collisions at RHIC and the LHC. In particular, we investigate in detail how these observables depend on different model ingredients, such as the initial heavy quark spectrum, the starting time of heavy-quark-medium interaction, the temperature profile of the medium in the pre-equilibrium stage, the energy loss and hadronization mechanisms of heavy quarks, the flow of medium, and the temperature dependence of heavy quark transport coefficient.

In this work, we focus on two most common heavy flavor observables: the nuclear modification factor $R_\mathrm{AA}$ and the elliptic flow coefficient $v_2$, which quantify the overall energy loss of heavy quarks and the asymmetry of heavy quark energy loss in different directions of the fireball. They are defined as follows:
\begin{align}
&R_\mathrm{AA}(p_\mathrm{T}) = \frac{1}{ N_\mathrm{coll}}\frac{dN^\mathrm{AA}/dp_\mathrm{T}}{dN^\mathrm{pp}/dp_\mathrm{T}},\\
&v_2(p_\mathrm{T}) = \langle \cos(2\phi)\rangle  = \left\langle \frac{p^{2}_{x} - p^{2}_{y}}{p^{2}_{x} + p^{2}_{y}}\right\rangle,
\end{align}
where $ N_\mathrm{coll}$ is the average number of binary collisions in a given centrality bin of AA collisions, and $\langle \ldots \rangle$ denotes the average over the final-state charmed hadrons observed in our simulations. In this work, we employ smooth hydrodynamic profiles. Therefore, $x$-$z$ defines the event plane and $x$-$y$ defines the transverse plane of AA collisions. Effects of event-by-event fluctuations on heavy flavor observables have been studied in our earlier work~\cite{Cao:2014fna,Cao:2017umt}, and found to be small.

With the above setups, we calculate $D$ meson $R_\mathrm{AA}$ and $v_{2}$ in 10-40\% Au-Au collisions at $\sqrt{s_\mathrm{NN}}=200$~GeV and 30-50\% Pb-Pb collisions at $\sqrt{s_\mathrm{NN}}=5.02$~TeV, and compare to STAR data at RHIC~\cite{Adam:2018inb,Acharya:2017qps} and ALICE and CMS data at the LHC~\cite{Acharya:2018hre,Acharya:2017qps,Sirunyan:2017plt}.

\begin{figure}[tbp]
    \centering
    \includegraphics[clip=,width=0.25\textwidth]{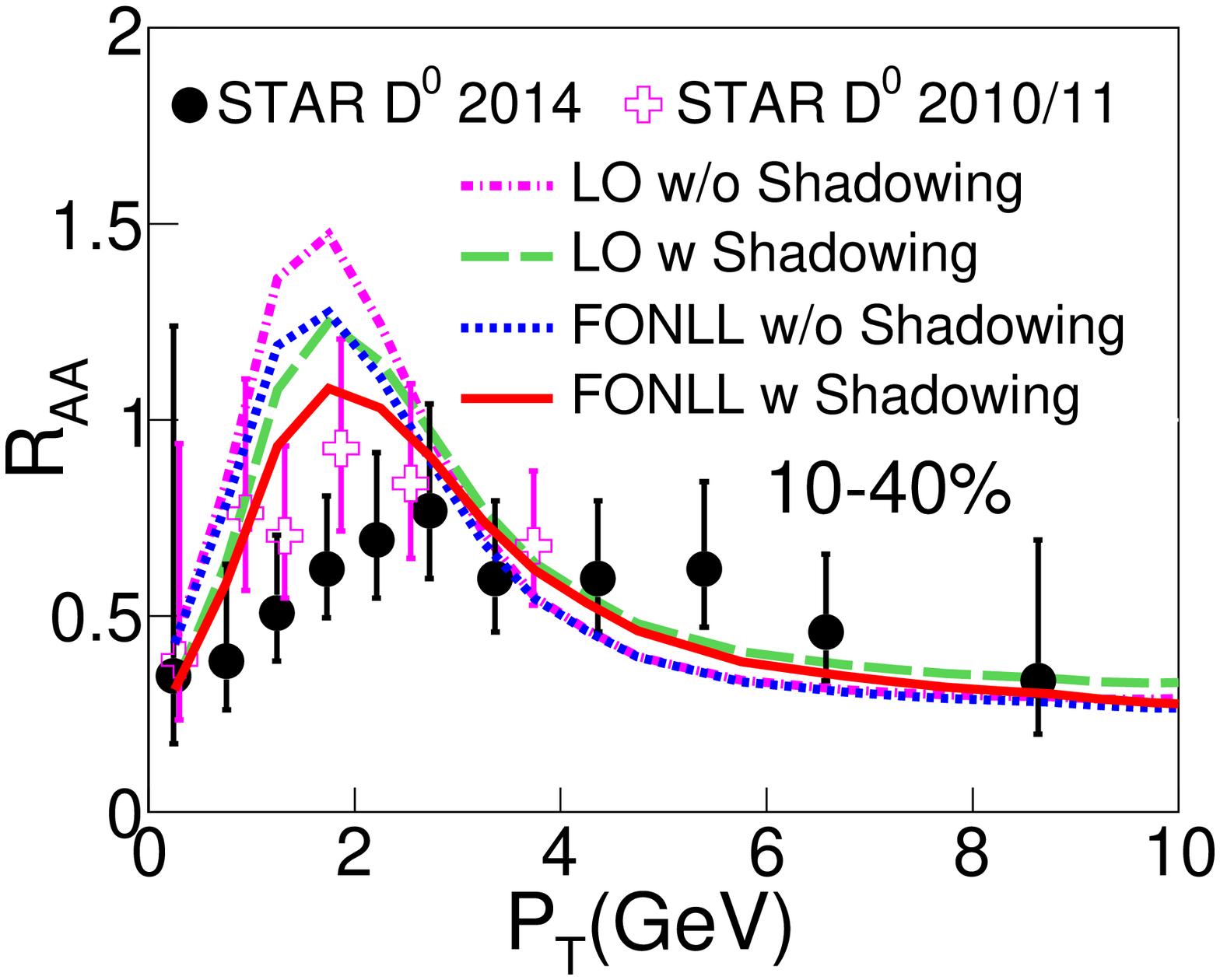}
    \hspace{-15pt}
    \includegraphics[clip=,width=0.25\textwidth]{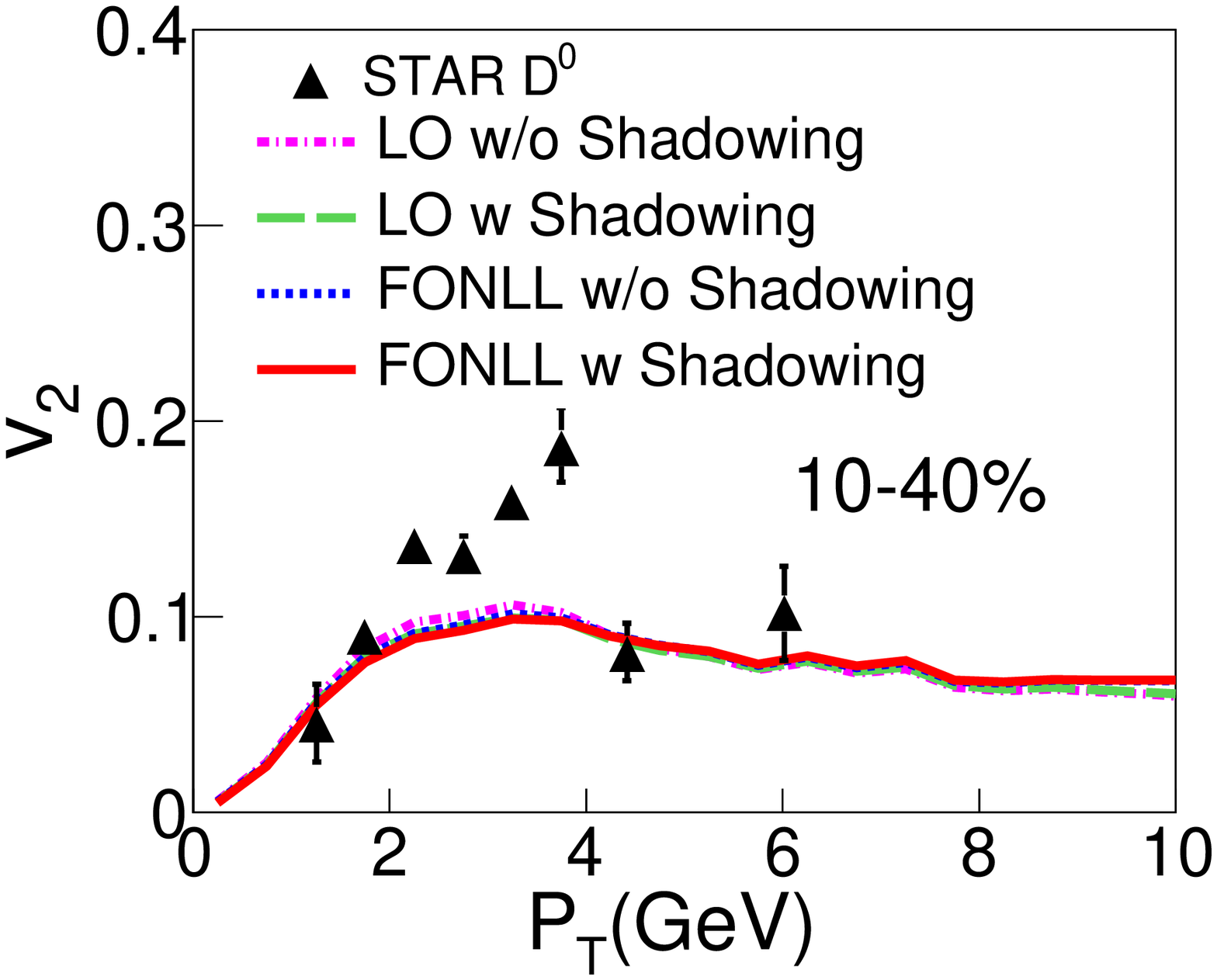}
    \includegraphics[clip=,width=0.25\textwidth]{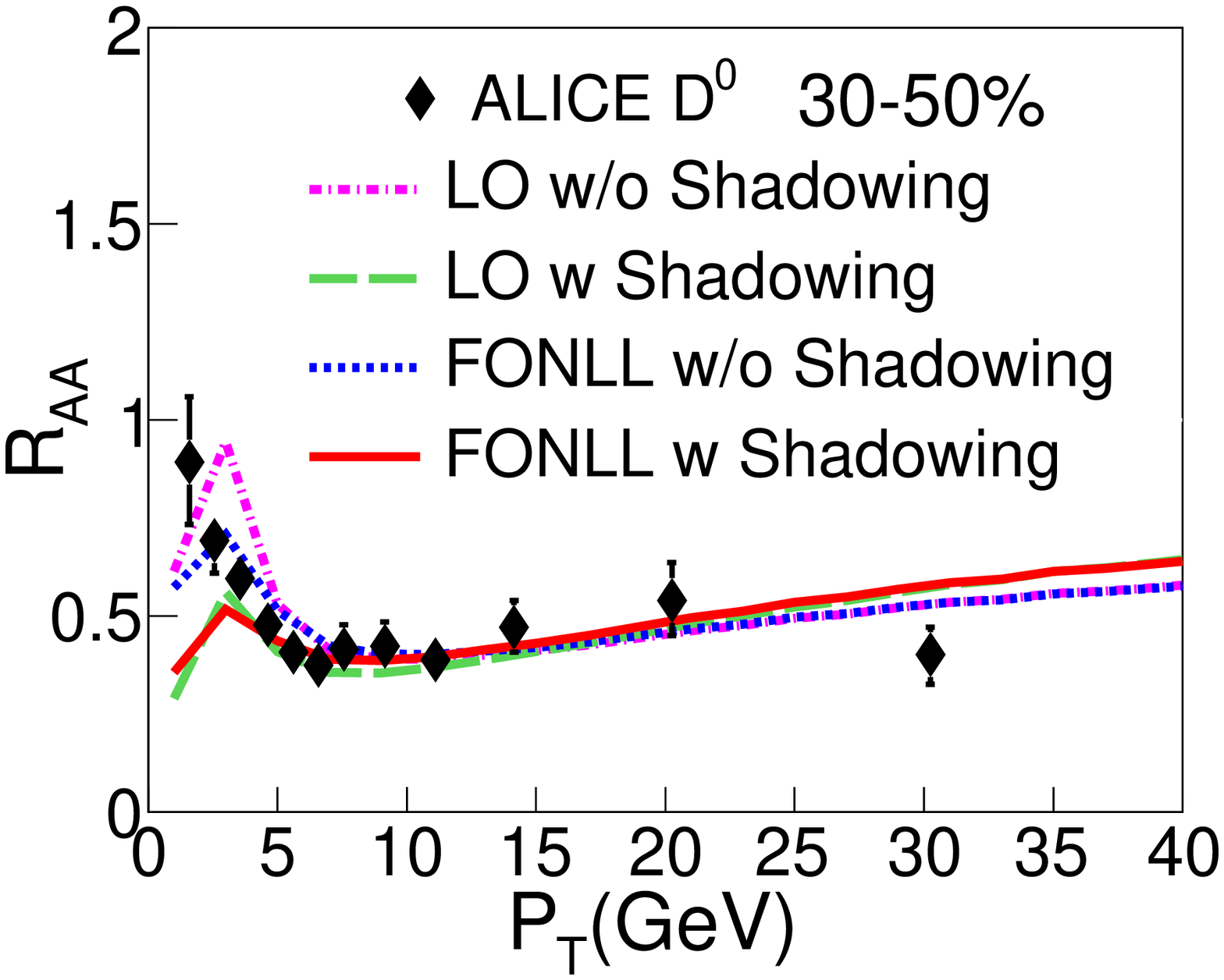}
    \hspace{-15pt}
    \includegraphics[clip=,width=0.25\textwidth]{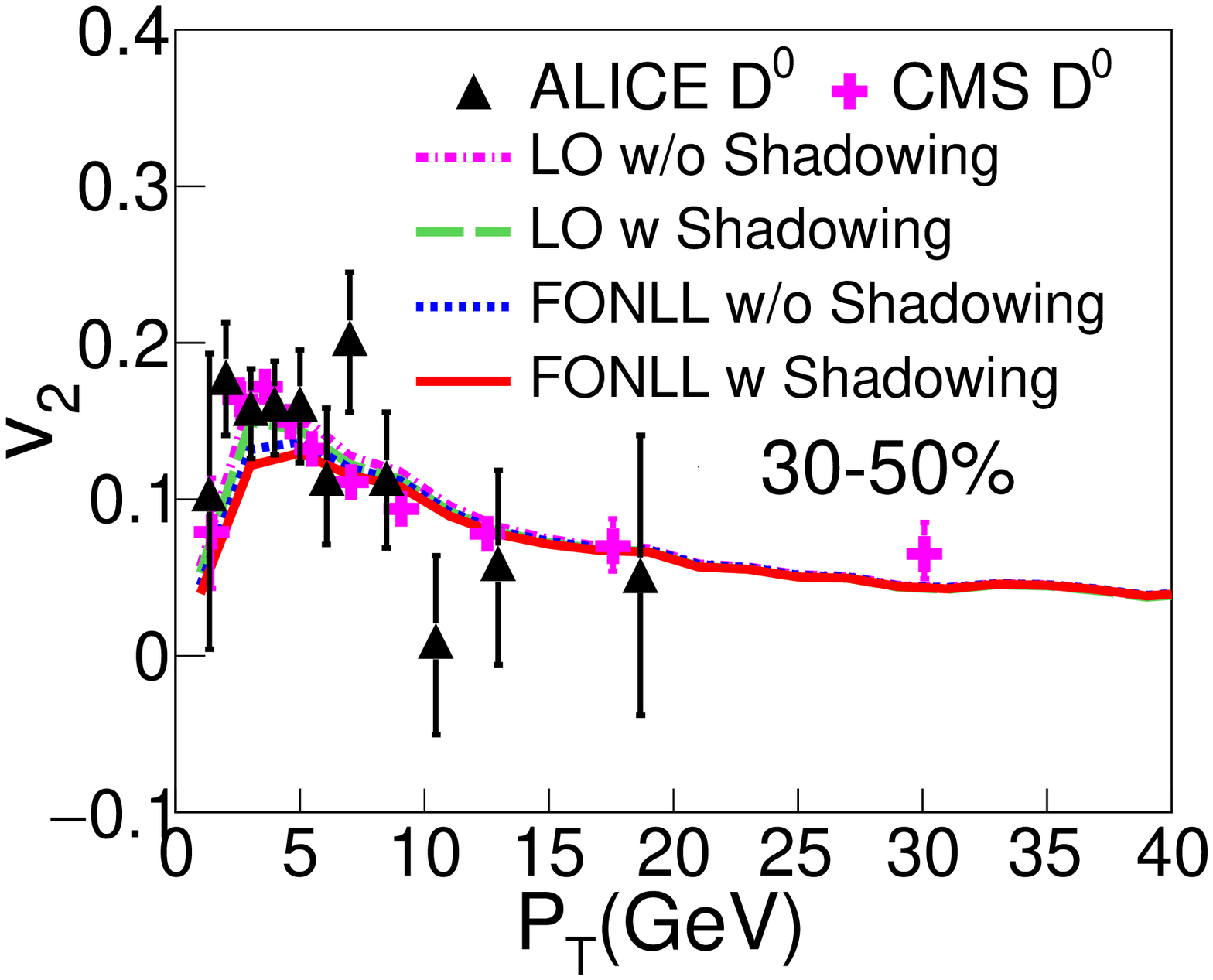}
    \caption{(Color online) Effects of initial heavy quark spectrum on $D$ meson $R_\mathrm{AA}$ (left panels) and $v_2$ (right panels) at RHIC (upper panels) and the LHC (lower panels).}
    \label{fig:initialSpectra}
\end{figure}

First, we investigate how $D$ meson $R_\mathrm{AA}$ and $v_2$ are affected by the initial heavy quark spectrum. In Fig.~\ref{fig:initialSpectra}, we compare the results for four different setups: using the LO spectra with and without nuclear shadowing effect, and using the FONLL spectra with and without nuclear shadowing effect. The diffusion coefficient of heavy quarks is set as $D_\mathrm{s}(2\pi T)=3$ at RHIC and 4 at the LHC in order to reasonably describe the experimental data on $R_\mathrm{AA}$. The smaller diffusion coefficient at RHIC than at the LHC can be understood as the stronger average jet-medium interaction at RHIC due to its lower average medium temperature~\cite{Burke:2013yra}. As mentioned earlier, the starting time of heavy-quark-medium interaction is set as $\tau_0=0.6$~fm for this comparison.

As shown in the left panels of Fig.~\ref{fig:initialSpectra}, without the nuclear shadowing effect, using the FONLL spectrum leads to smaller $R_\mathrm{AA}$ at low $p_\mathrm{T}$ than using the LO spectrum (13\% smaller at RHIC and 25\% smaller at the LHC), while no apparent difference is observed at high $p_\mathrm{T}$. This is consistent with the findings in Ref.~\cite{Rapp:2018qla} since the FONLL spectrum is lower than the LO spectrum below $p_\mathrm{T}\sim 5$~GeV, but they become similar above $5$~GeV. The inclusion of nuclear shadowing effect significantly suppresses the initial spectrum of charm quarks in AA collisions at low $p_\mathrm{T}$ compared to that in pp collisions (15\% at RHIC and 27\% at the LHC when the EPPS16 parameterization is applied within the FONLL framework), while at high $p_\mathrm{T}$ the spectrum is slightly enhanced (anti-shadowing). This effect is directly reflected in the result of $D$ meson $R_\mathrm{AA}$. For elliptic flow $v_2$, we can see from the right panels of Fig.~\ref{fig:initialSpectra} that different spectra yield negligible effect on $D$ meson $v_2$ at RHIC. At the LHC, using the FONLL spectrum yields about 19\% smaller $v_2$ than using the LO spectrum below $p_\mathrm{T}\sim 5$~GeV. For the rest of our work, we will use the FONLL spectrum combined with the EPPS16 parametrization of nuclear shadowing in our calculations.

\begin{figure}[tbp]
    \centering
    \includegraphics[clip=,width=0.25\textwidth]{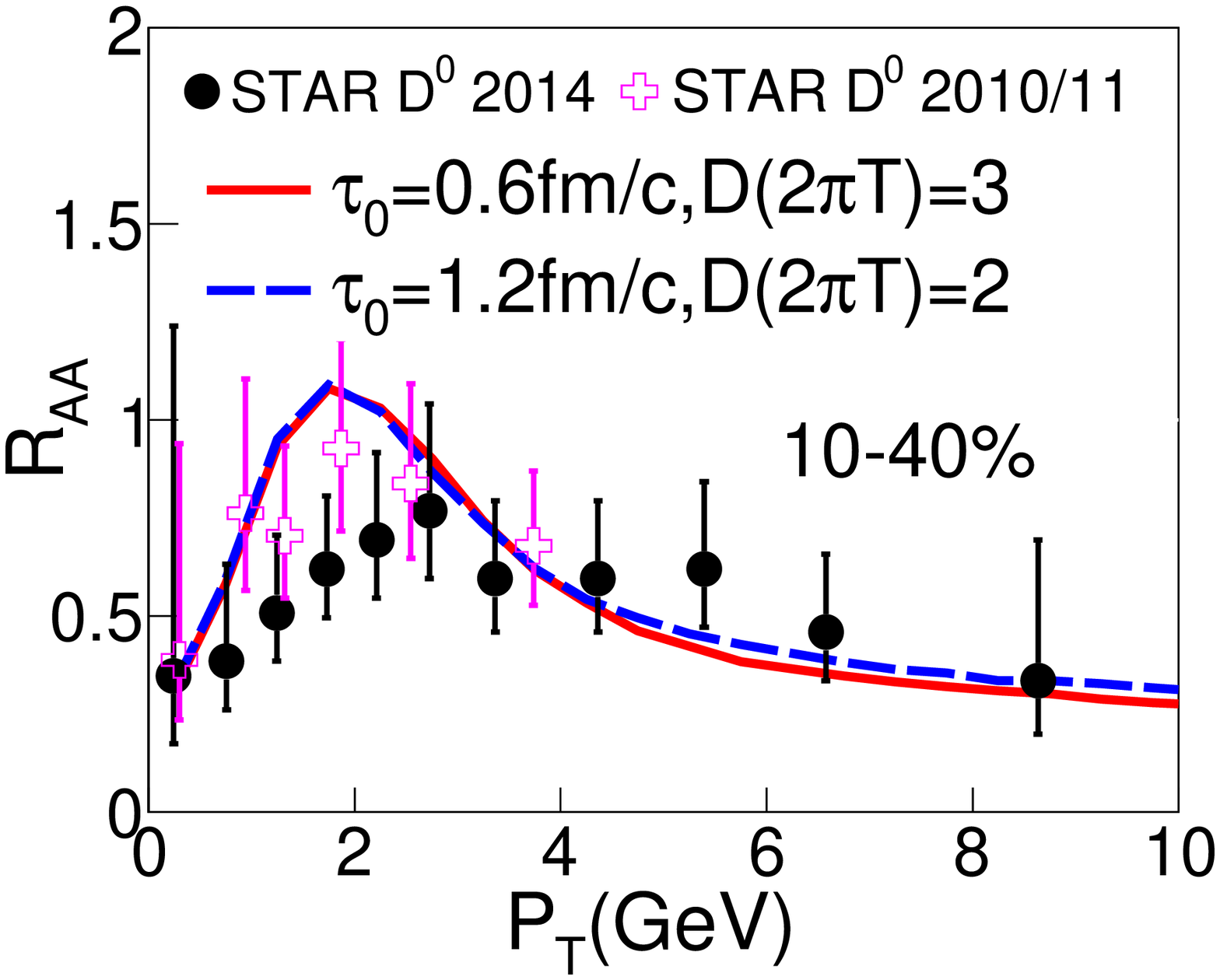}
    \hspace{-15pt}
    \includegraphics[clip=,width=0.25\textwidth]{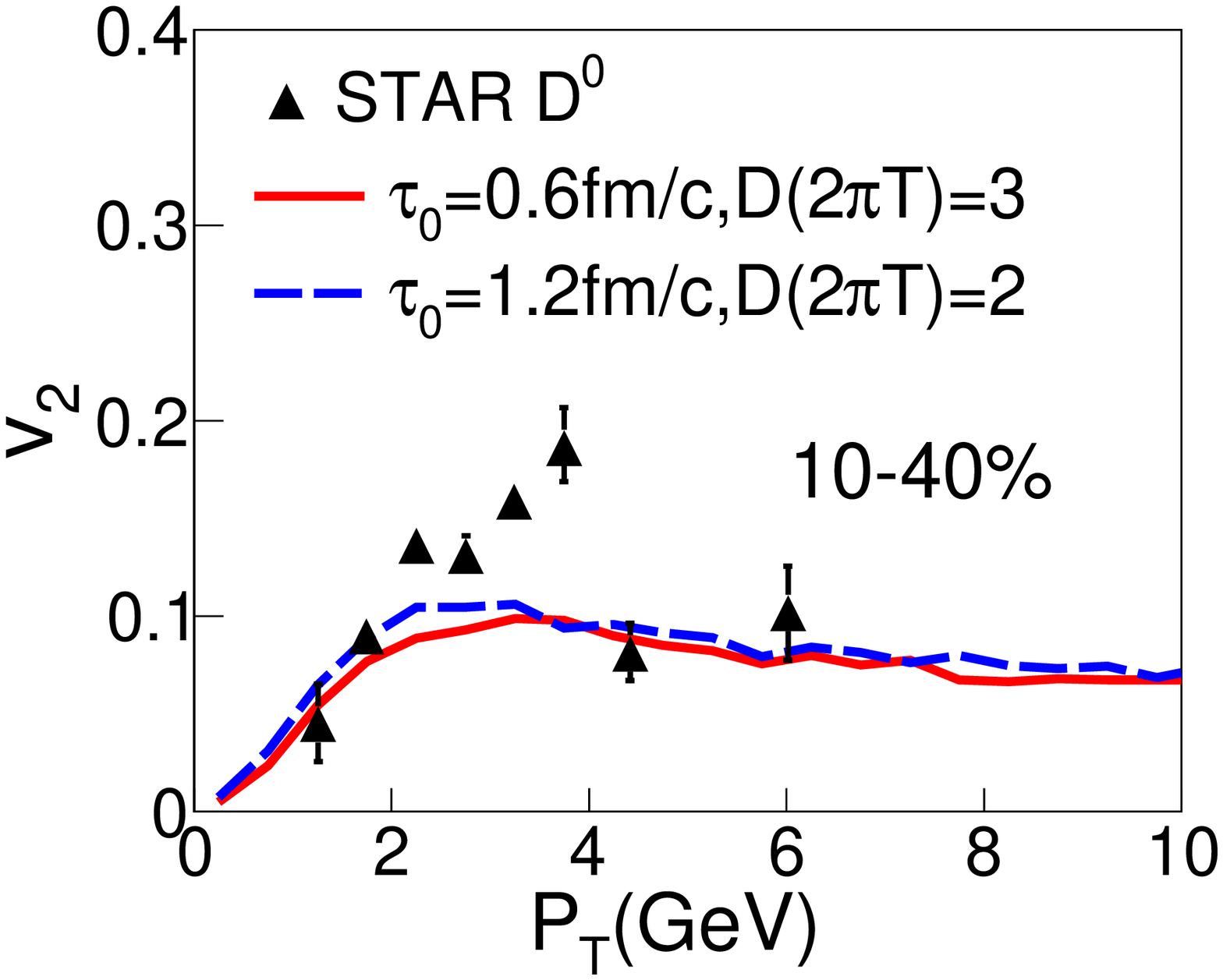}
    \includegraphics[clip=,width=0.25\textwidth]{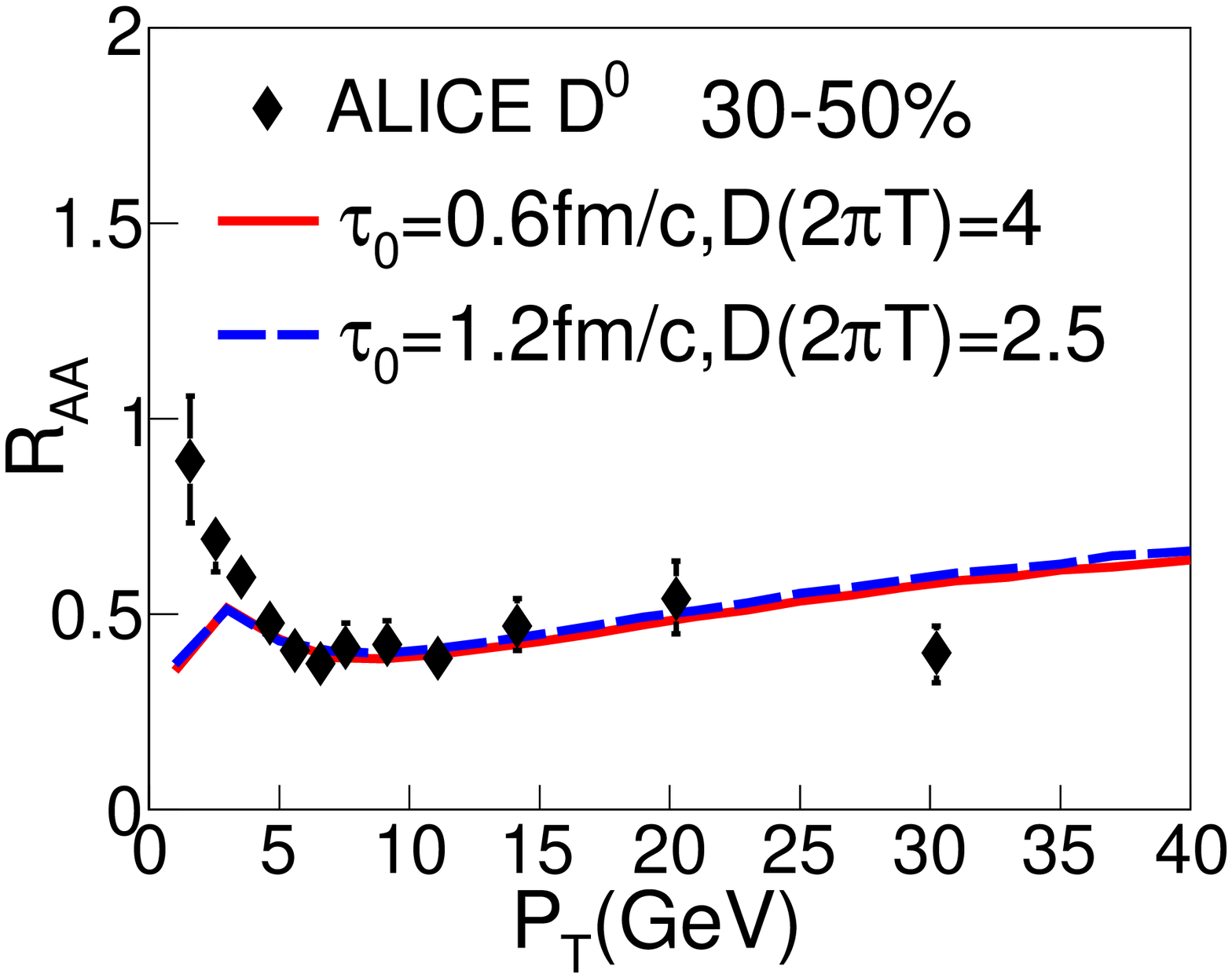}
    \hspace{-15pt}
    \includegraphics[clip=,width=0.25\textwidth]{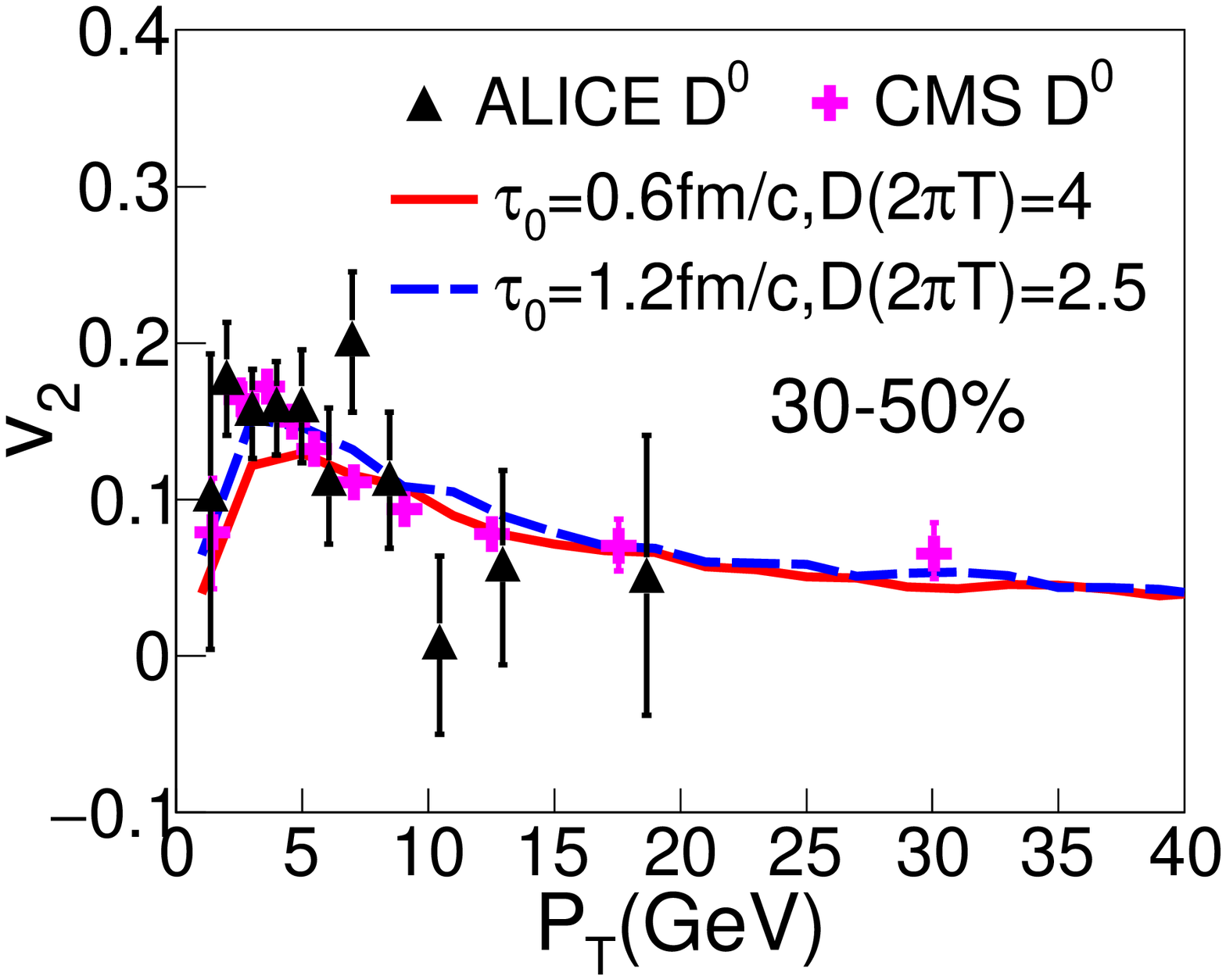}
    \caption{(Color online) Effects of the starting time $\tau_0$ of heavy-quark-medium interactions on $D$ meson $R_\mathrm{AA}$ (left panels) and $v_2$ (right panels) at RHIC (upper panels) and the LHC (lower panels).}
    \label{fig:initialTime}
\end{figure}

To date, we know little about jet-medium interaction in the pre-equilibrium stage of heavy-ion collisions. In most literature, energetic particles are assumed to stream freely in this stage until the thermalized QGP forms (e.g. $\tau _{0}=0.6$~fm). However, as shown in Refs.~\cite{Mrowczynski:2017kso,Carrington:2020sww}, interactions within this stage could be strong. Therefore, it is worth quantifying the possible uncertainties in jet quenching observables from different model assumptions of the pre-equilibrium stage.

In Fig.~\ref{fig:initialTime}, we first compare two different choices of the starting time $\tau_0$ of heavy-quark-medium interaction (0.6~fm {\it vs.} 1.2~fm). To look for the pure effect of different starting times on $D$ meson $v_2$, we first adjust the diffusion coefficient such that the two different choices of $\tau_0$ yield similar $D$ meson $R_\mathrm{AA}$, as shown in the left panels of Fig.~\ref{fig:initialTime}. Due to the shorter evolution time of heavy quarks inside the QGP, using a later starting time for heavy-quark-medium interaction (i.e., $\tau_0=1.2$~fm) requires about 35\% smaller diffusion coefficient $D_\mathrm{s}$, which means stronger coupling strength between heavy quarks and QGP, in order to reproduce the same suppression as using the starting time of $\tau_0=0.6$~fm. With the same $D$ meson $R_\mathrm{AA}$, a larger $D$ meson $v_2$ is observed at low $p_\mathrm{T}$ in the right panels for $\tau_0=1.2$~fm compared to $\tau_0=0.6$ (8\% larger at RHIC and 24\% at the LHC). This is because shifting more interaction towards later time of the QGP evolution, when the anisotropic flow of medium is stronger, allows low $p_\mathrm{T}$ (near-thermal) heavy quarks to pick up more $v_2$ from the nuclear medium.

\begin{figure}[tbp]
    \centering
    \includegraphics[clip=,width=0.25\textwidth]{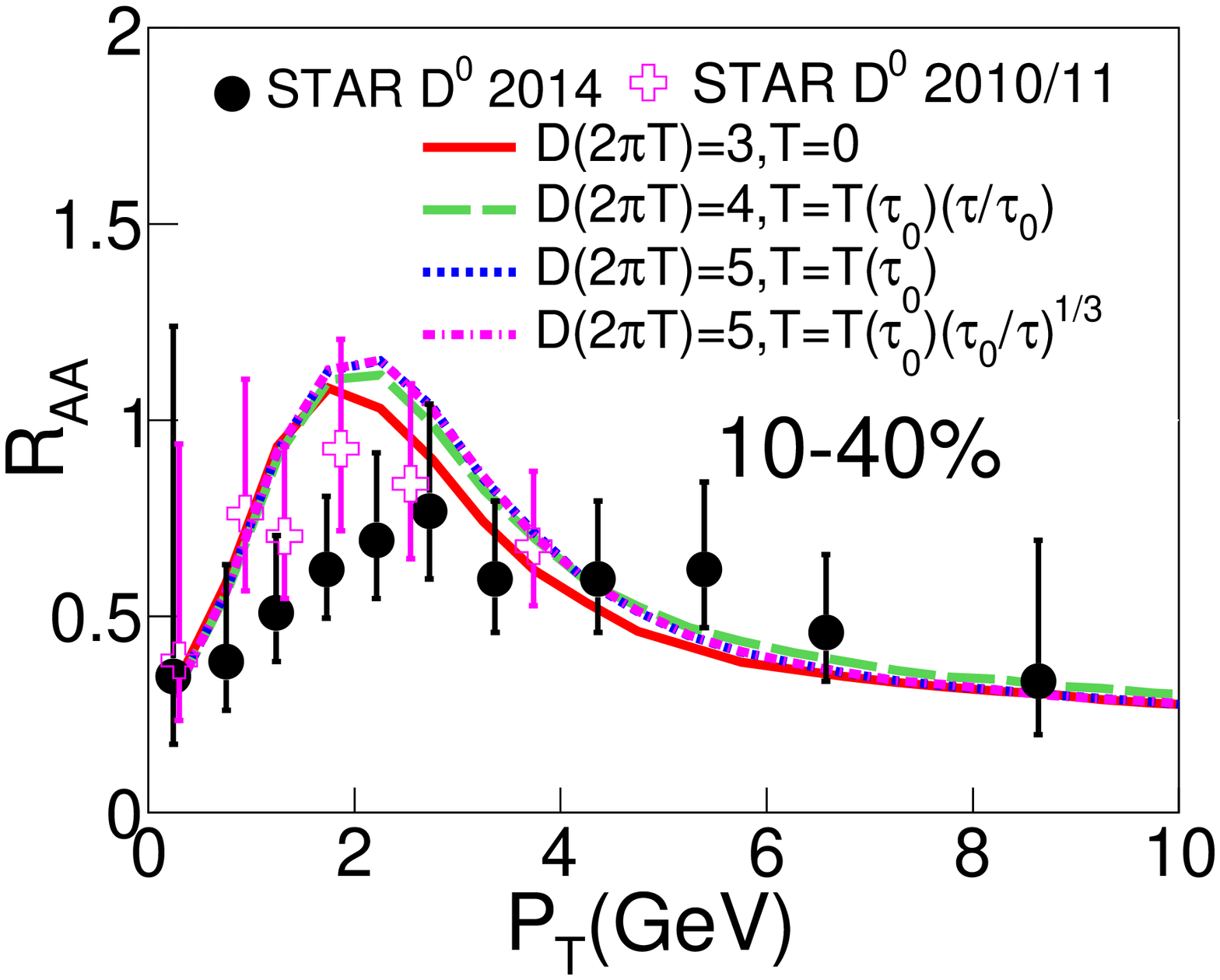}
    \hspace{-15pt}
    \includegraphics[clip=,width=0.25\textwidth]{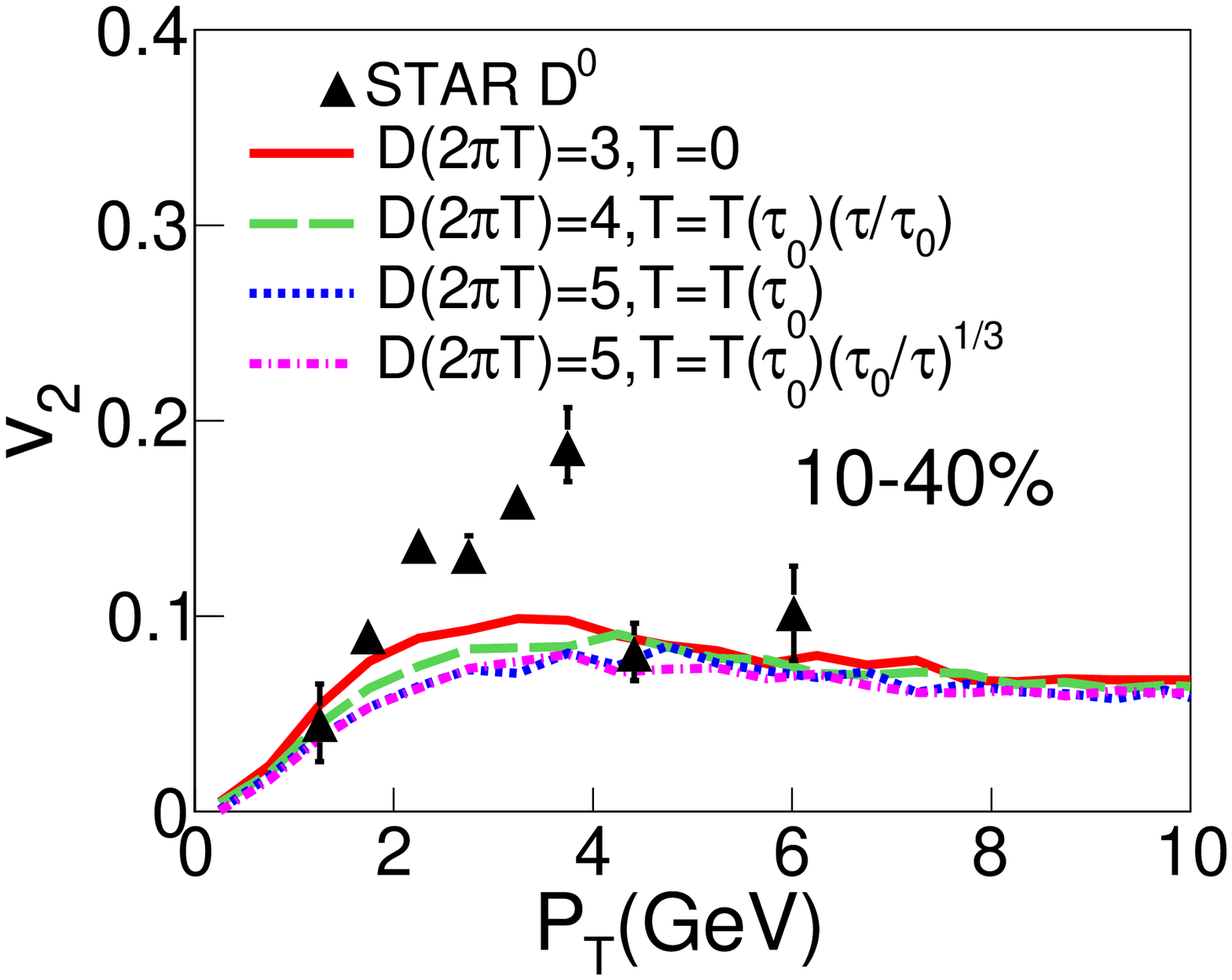}
    \includegraphics[clip=,width=0.25\textwidth]{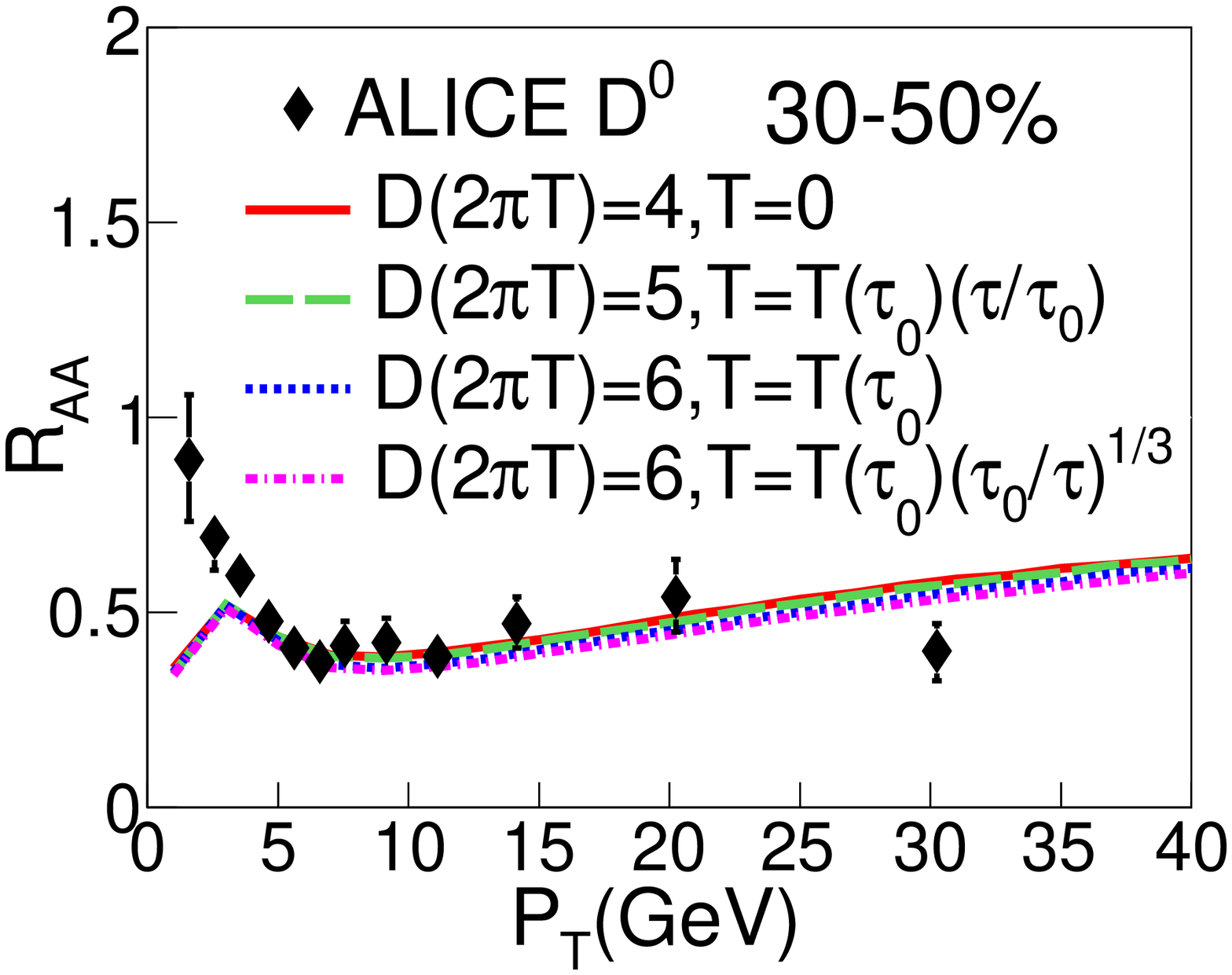}
    \hspace{-15pt}
    \includegraphics[clip=,width=0.25\textwidth]{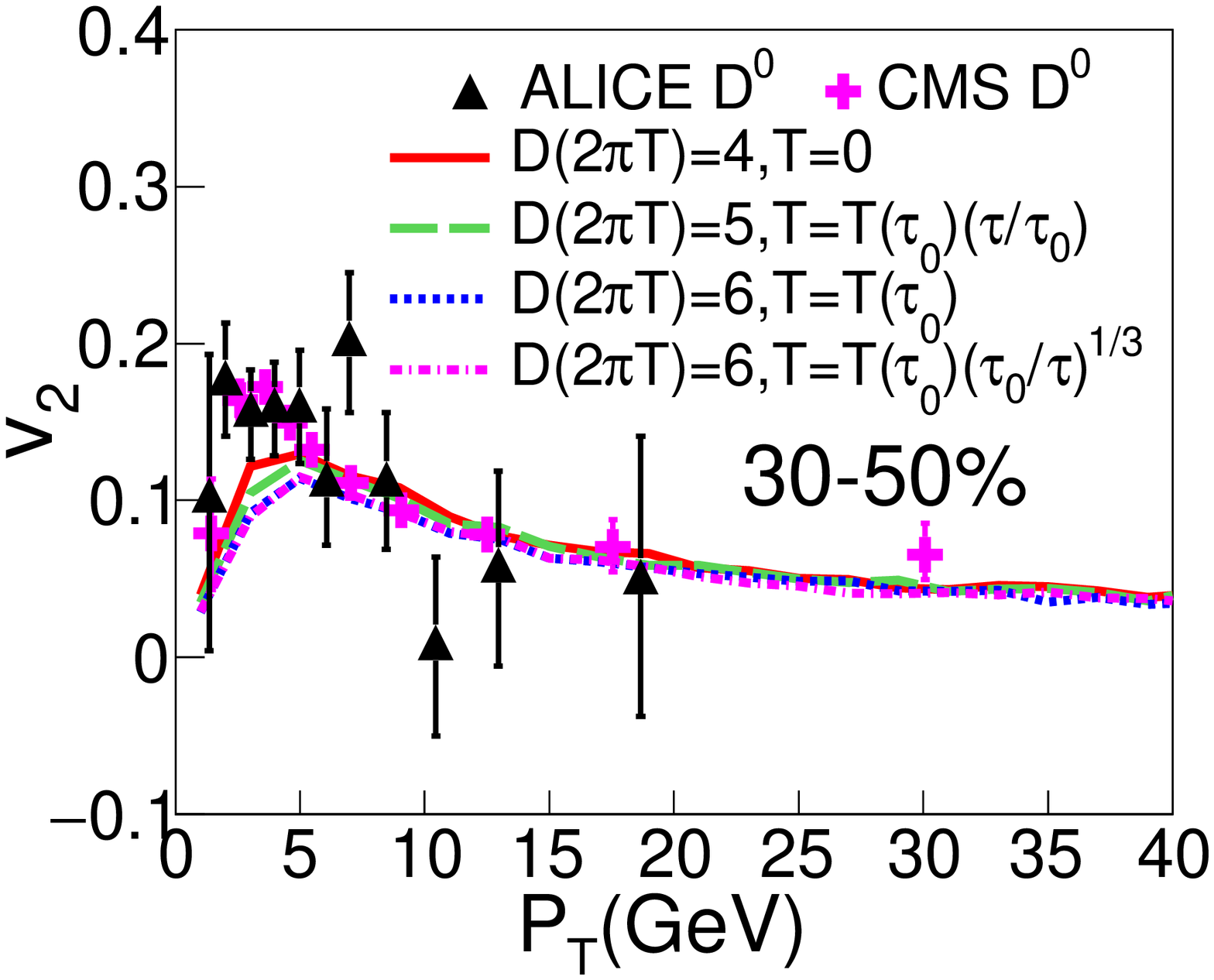}
    \caption{(Color online) Effects of different temperature profiles of the pre-equilibrium stage on $D$ meson $R_\mathrm{AA}$ (left panels) and $v_2$ (right panels) at RHIC (upper panels) and the LHC (lower panels).}
    \label{fig:initialTemperature}
\end{figure}

To further investigate how different assumptions of heavy-quark-medium interaction in the pre-equilibrium stage affect heavy quark observables, we extend the above study beyond the free-streaming hypothesis for heavy quarks before $\tau_0$. We compare different modelings of the temperature profiles of the pre-equilibrium stage before the thermalized QGP forms. For heavy quark evolution in the pre-equilibrium stage ($0<\tau<\tau_0$), we still utilize the modified Langevin dynamics [Eq.~(\ref{eq:Langevin})] in the same way as in the QGP phase ($\tau>\tau_0$) as long as the temperature profile of the background medium is provided. As illustrated in Fig.~\ref{fig:initialTemperature}, four different scenarios are used for the temperature profile of the medium before $\tau_0$ (0.6~fm): (1) $T(\tau)=0$ -- free-streaming, (2) $T(\tau)=T(\tau _{0})(\tau/\tau_0)$ -- a linear increase from 0 to $\tau_0$, (3) $T(\tau) = T(\tau_0)$ -- constant temperature before $\tau_0$, and (4) $T(\tau)=T(\tau _{0})(\tau_0/\tau)^{1/3}$ -- Bjorken evolution profiles before $\tau_0$.

Similar to the previous study in Fig.~\ref{fig:initialTime}, we adjust the diffusion coefficient such that these different model setups provide similar $D$ meson $R_\mathrm{AA}$ in the left panels of Fig.~\ref{fig:initialTemperature}. One can see that a larger $D_\mathrm{s}$, or weaker heavy-quark-medium coupling strength, is required when a higher average medium temperature is modeled for the pre-equilibrium stage. After $D$ meson $R_\mathrm{AA}$ is fixed, we can see from the right panels that different assumptions of the pre-equilibrium temperature profiles give rise to different elliptic flow $v_2$ of $D$ mesons. The free-streaming assumption yields about 39\% (19\%) larger $v_2$ than the constant temperature and Bjorken evolution profiles at low $p_\mathrm{T}$ at RHIC (the LHC). The elliptic flow $v_2$ for the linear increasing temperature profile lies in the middle. Very little effect is observed in high $p_\mathrm{T}$ regime. This is consistent with the findings in Fig.~\ref{fig:initialTime}: stronger heavy-quark-medium interaction at a later time results in a larger $v_2$ (after tuning the model parameter to describe $R_\mathrm{AA}$). Our results are qualitatively consistent with the result in Ref.~\cite{Andres:2019eus}, though the effect found in our study is quantitatively smaller, especially at high $p_\mathrm{T}$. Keeping in mind the above uncertainties from the pre-equilibrium stage, we now turn to explore the effects from other model ingredients using the free-streaming hypothesis before $\tau_0=0.6$~fm in the rest of this work.

\begin{figure}[tbp]
    \centering
    \includegraphics[clip=,width=0.25\textwidth]{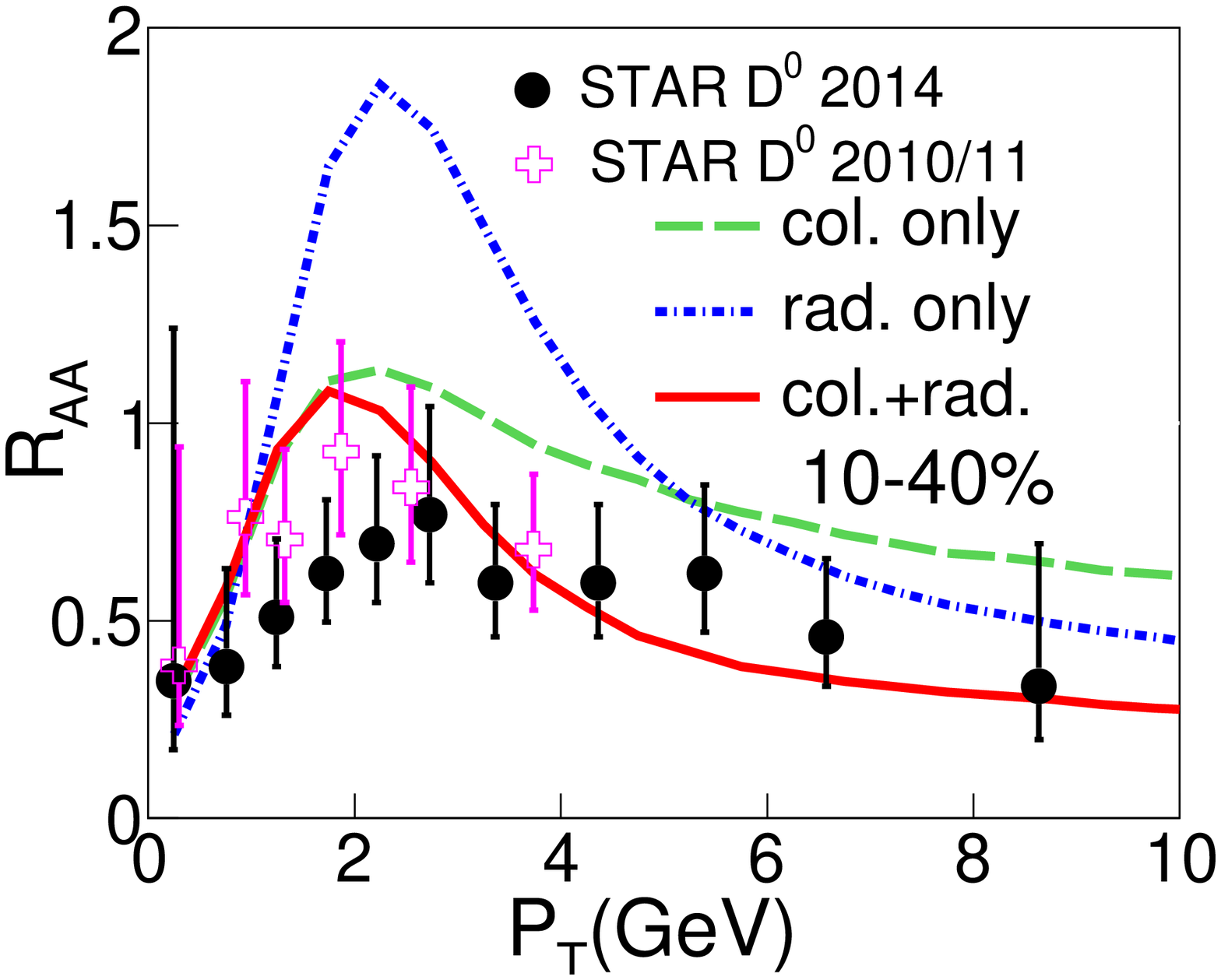}
    \hspace{-15pt}
    \includegraphics[clip=,width=0.25\textwidth]{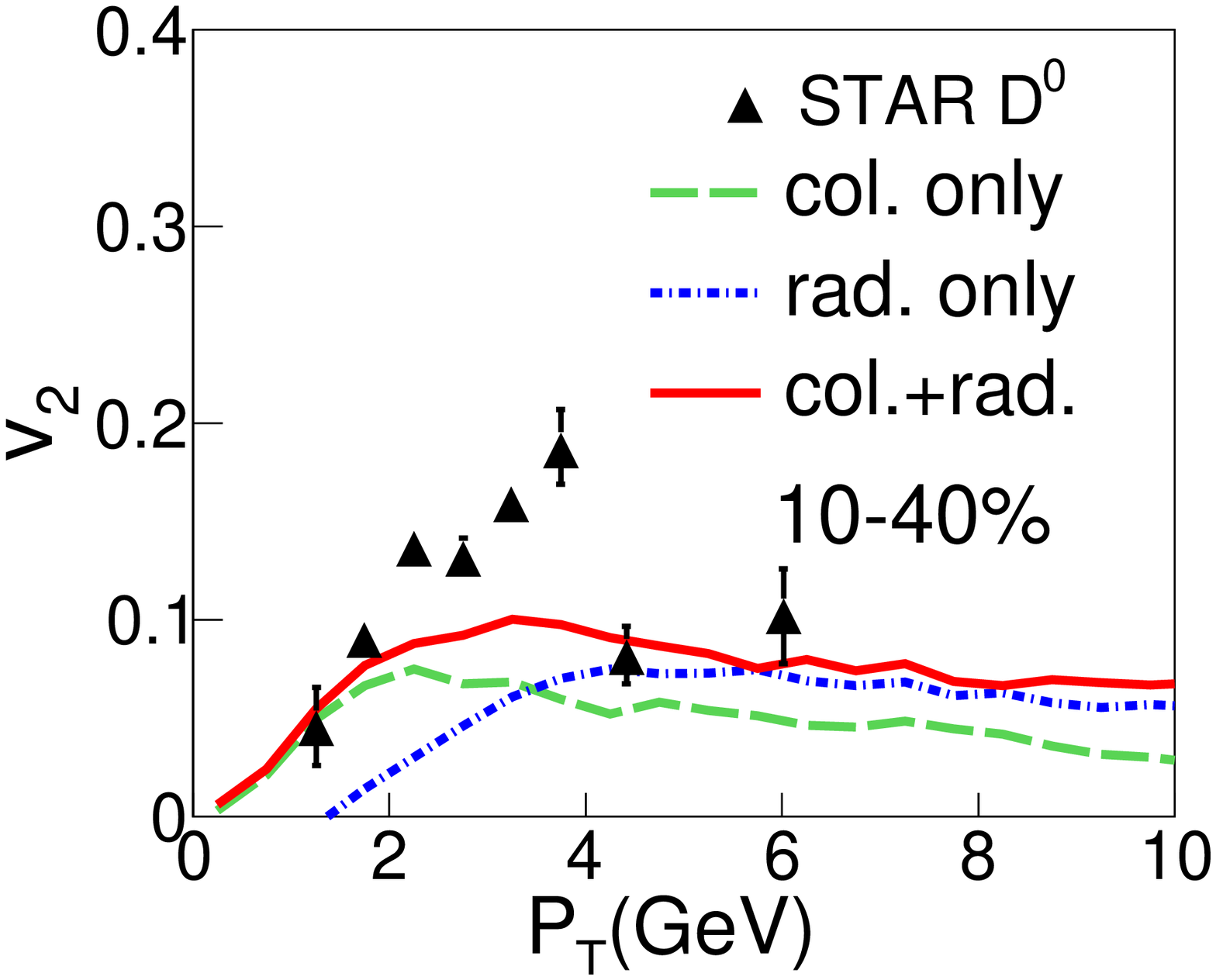}
    \includegraphics[clip=,width=0.25\textwidth]{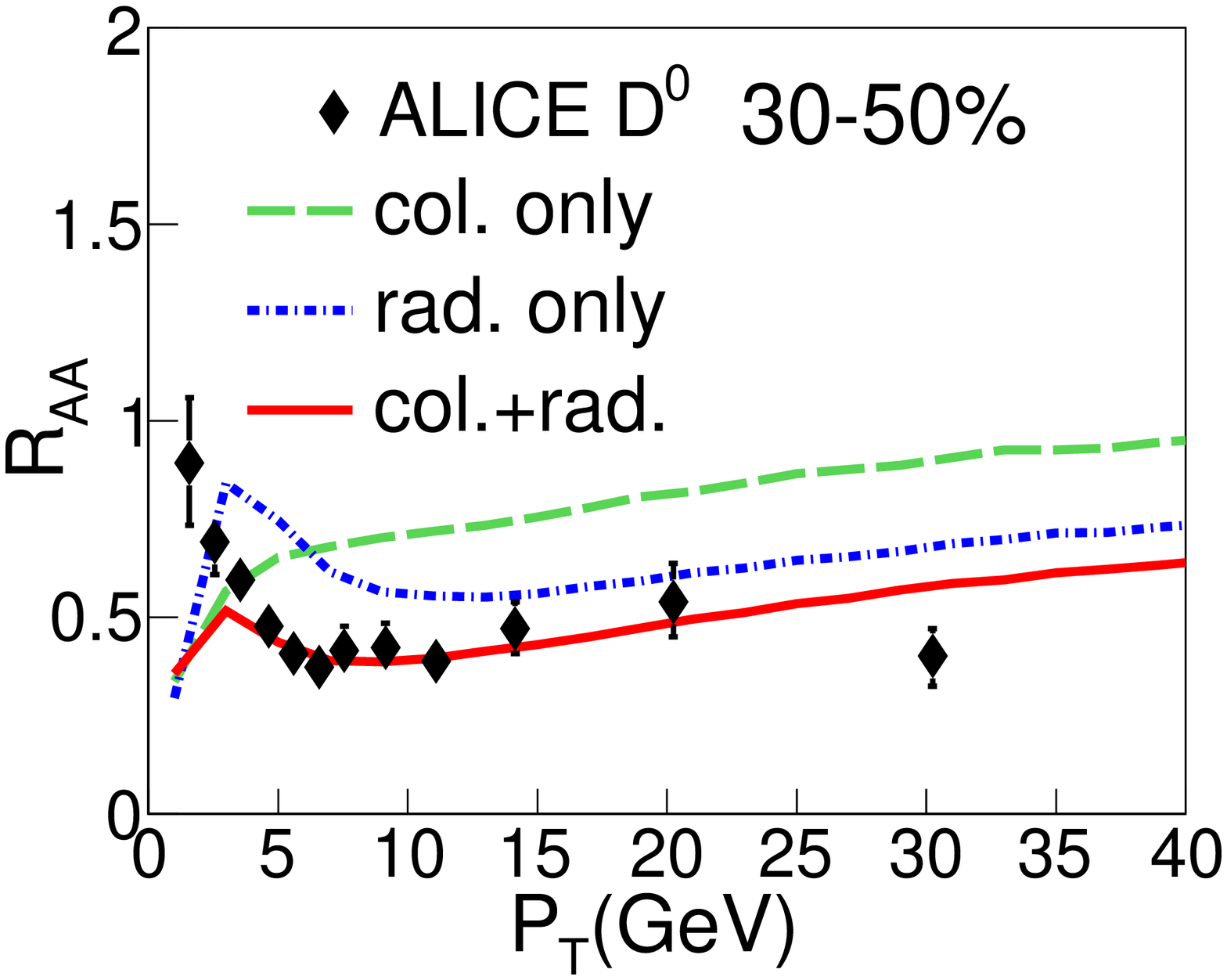}
    \hspace{-15pt}
    \includegraphics[clip=,width=0.25\textwidth]{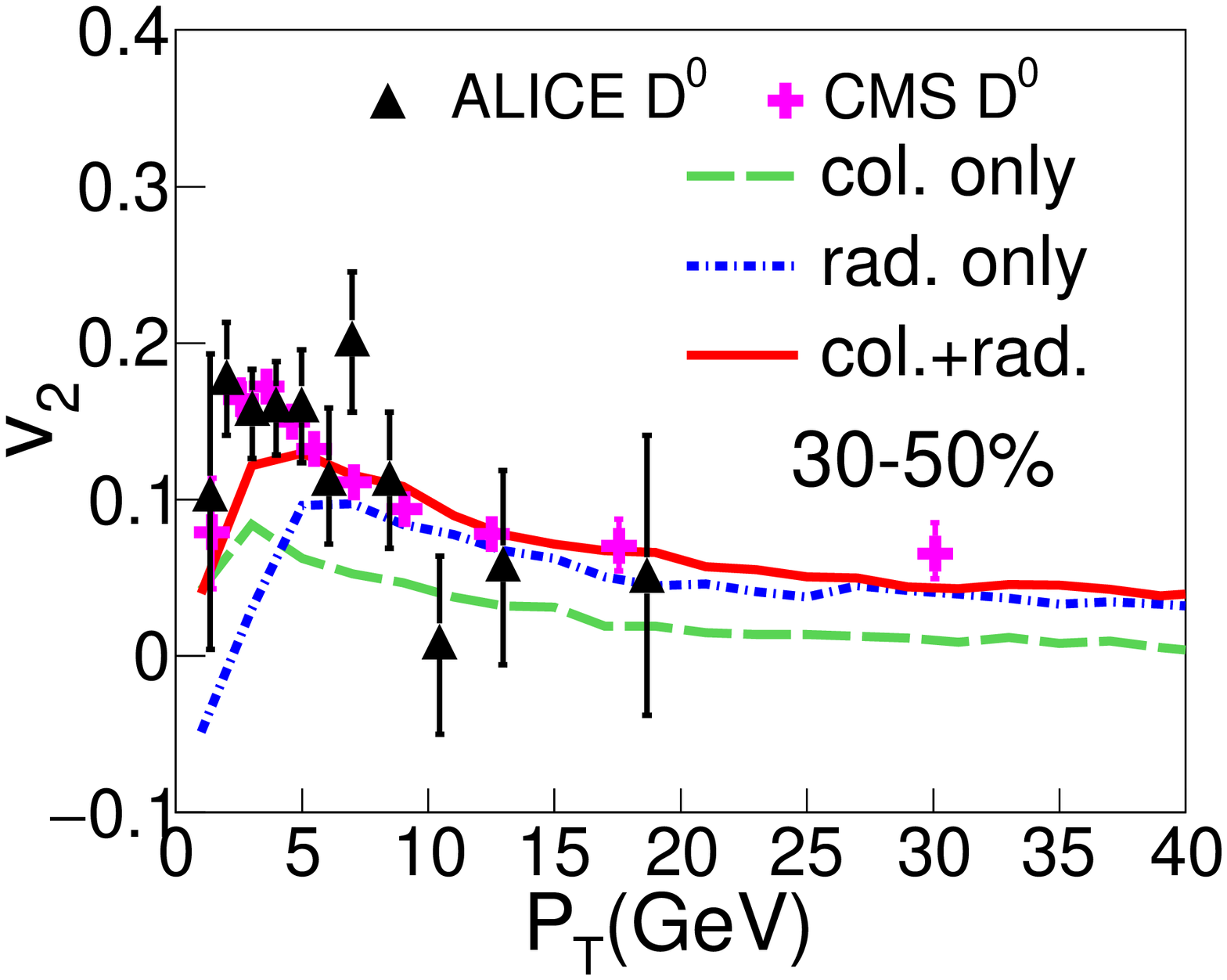}
    \caption{(Color online) Effects of collisional {\it vs.} radiative energy loss mechanisms on $D$ meson $R_\mathrm{AA}$ (left panels) and $v_2$ (right panels) at RHIC (upper panels) and the LHC (lower panels).}
    \label{fig:energyLoss}
\end{figure}

The energy loss of heavy quarks inside the QGP is the main cause for the nuclear modification of $D$ mesons in heavy-ion collisions. Therefore, it is crucial to understand the detailed energy loss mechanism for heavy quarks in QGP. In Fig.~\ref{fig:energyLoss}, we study how collisional and radiative energy losses contribute to $D$ meson $R_\mathrm{AA}$ and $v_2$, individually. The diffusion coefficient of charm quarks is set as $D_\mathrm{s}(2\pi T)=3$ at RHIC and 4 at the LHC such that our model provides a reasonable description of the experimental data after the inclusion of both collisional and radiative energy loss mechanisms.

From $R_\mathrm{AA}$ (left panels) and $v_2$ (right panels) in Fig.~\ref{fig:energyLoss}, we observe that collisional energy loss dominates heavy quark evolution at low $p_\mathrm{T}$ while radiative energy loss dominates at high $p_\mathrm{T}$. For $D$ meson $R_\mathrm{AA}$, the crossing point is around 5~GeV at RHIC and around 7~GeV at the LHC. The slight difference of this crossing point at RHIC and the LHC is mainly caused by different initial charm quark spectra and different QGP flow velocities at these two colliding energies. One can also see that neither collisional nor radiative mechanism alone is sufficient to describe the $p_\mathrm{T}$ dependence of $D$ meson observables.

\begin{figure}[tbp]
    \centering
    \includegraphics[clip=,width=0.25\textwidth]{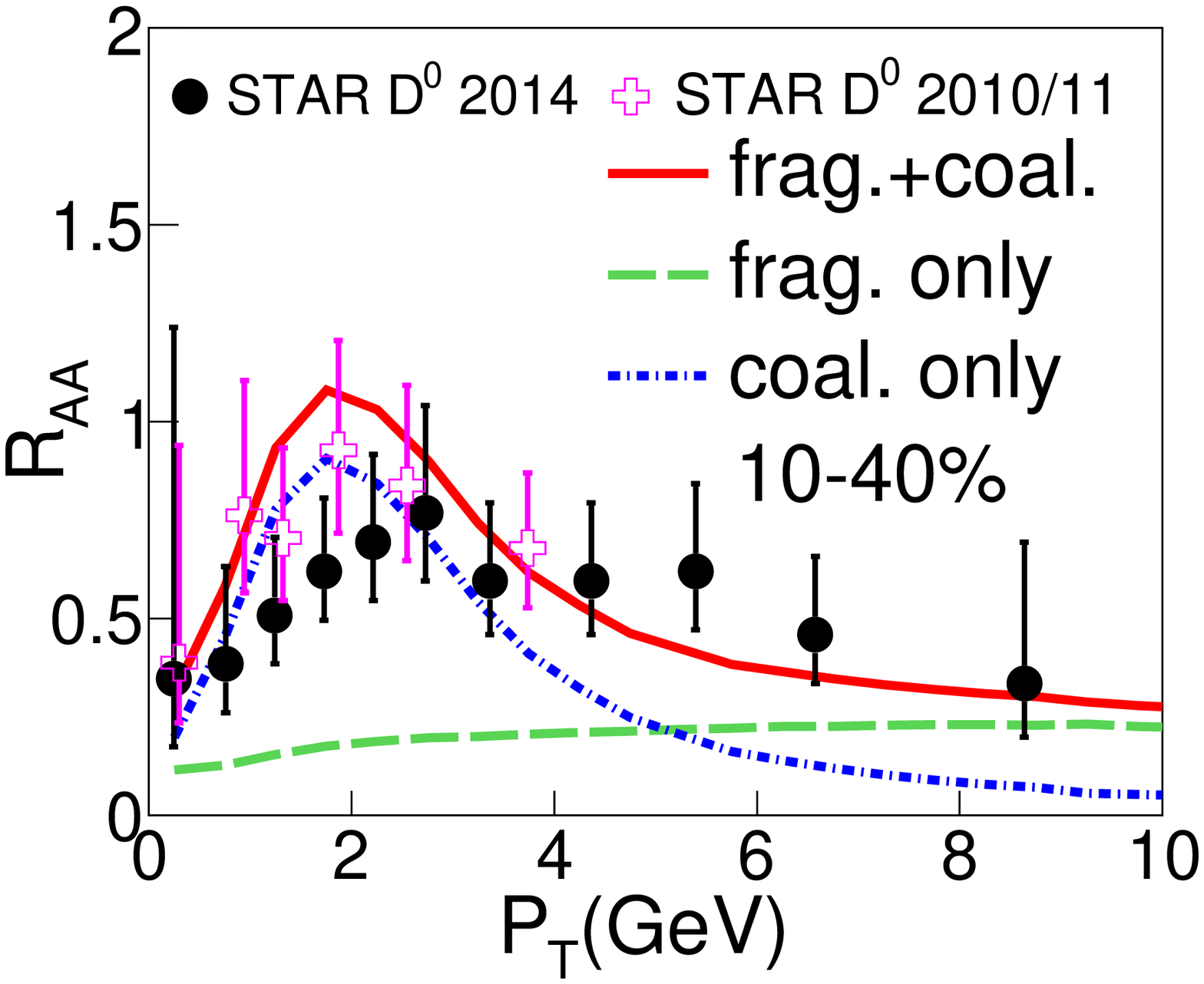}
    \hspace{-15pt}
    \includegraphics[clip=,width=0.25\textwidth]{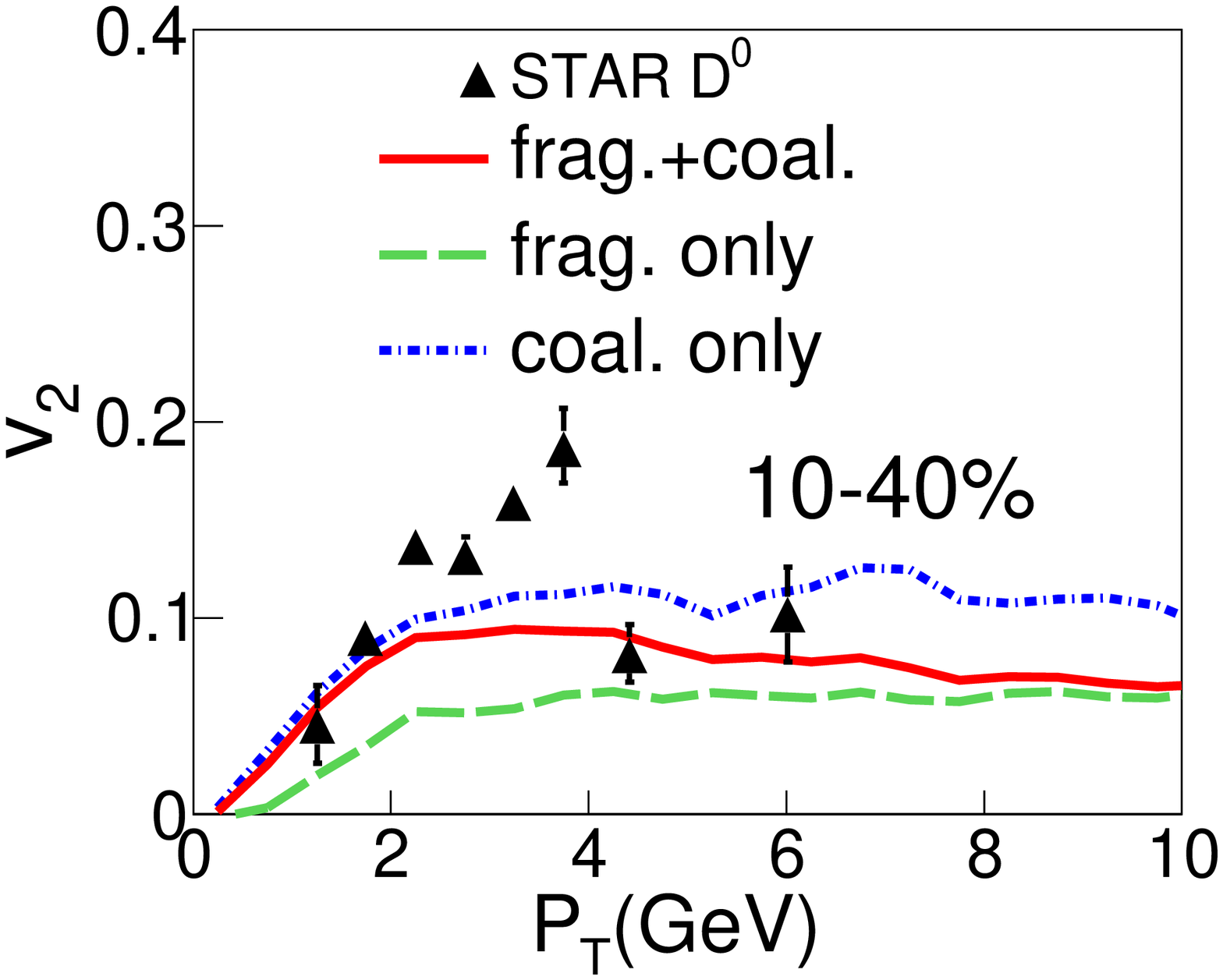}
    \includegraphics[clip=,width=0.25\textwidth]{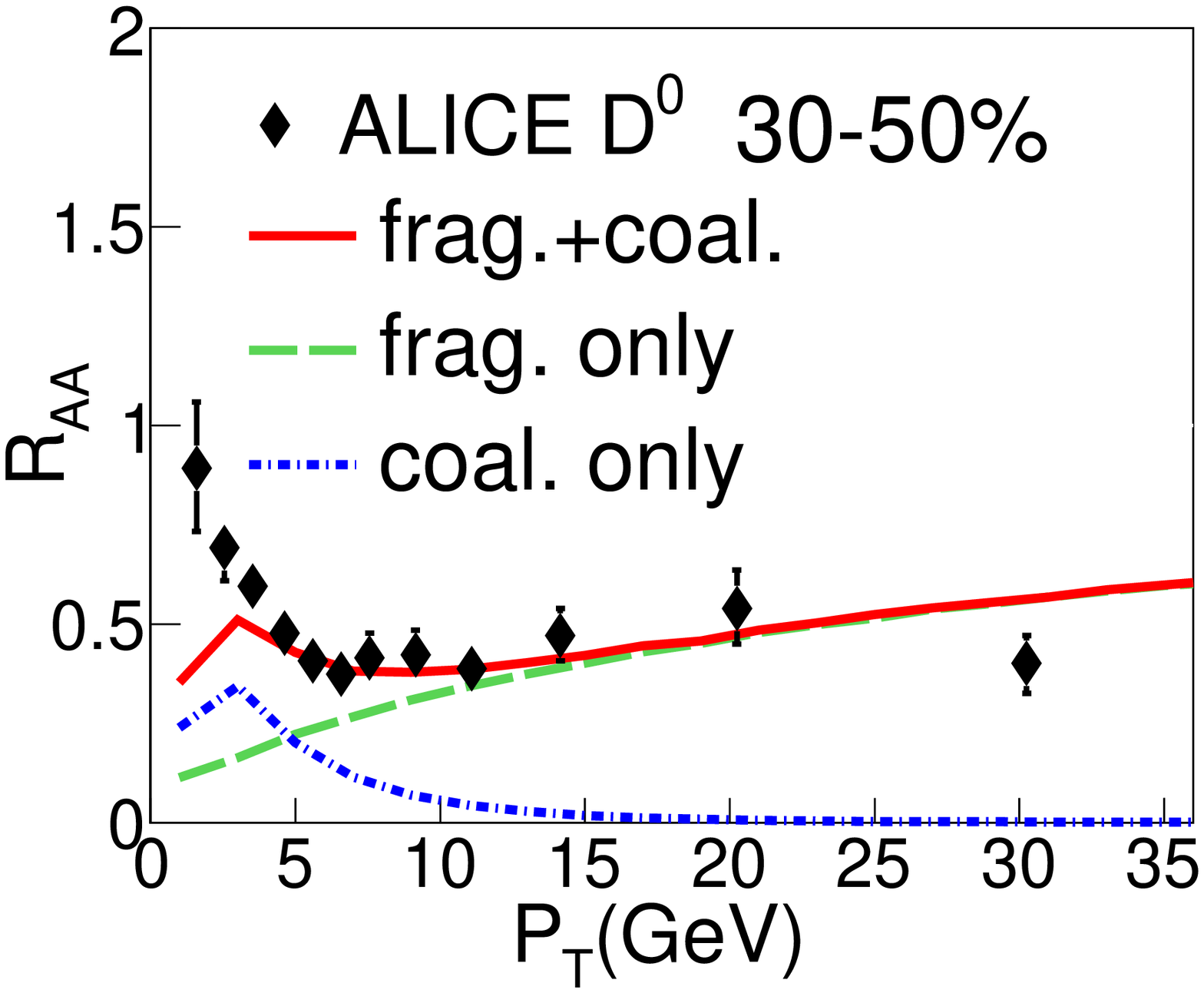}
    \hspace{-15pt}
    \includegraphics[clip=,width=0.25\textwidth]{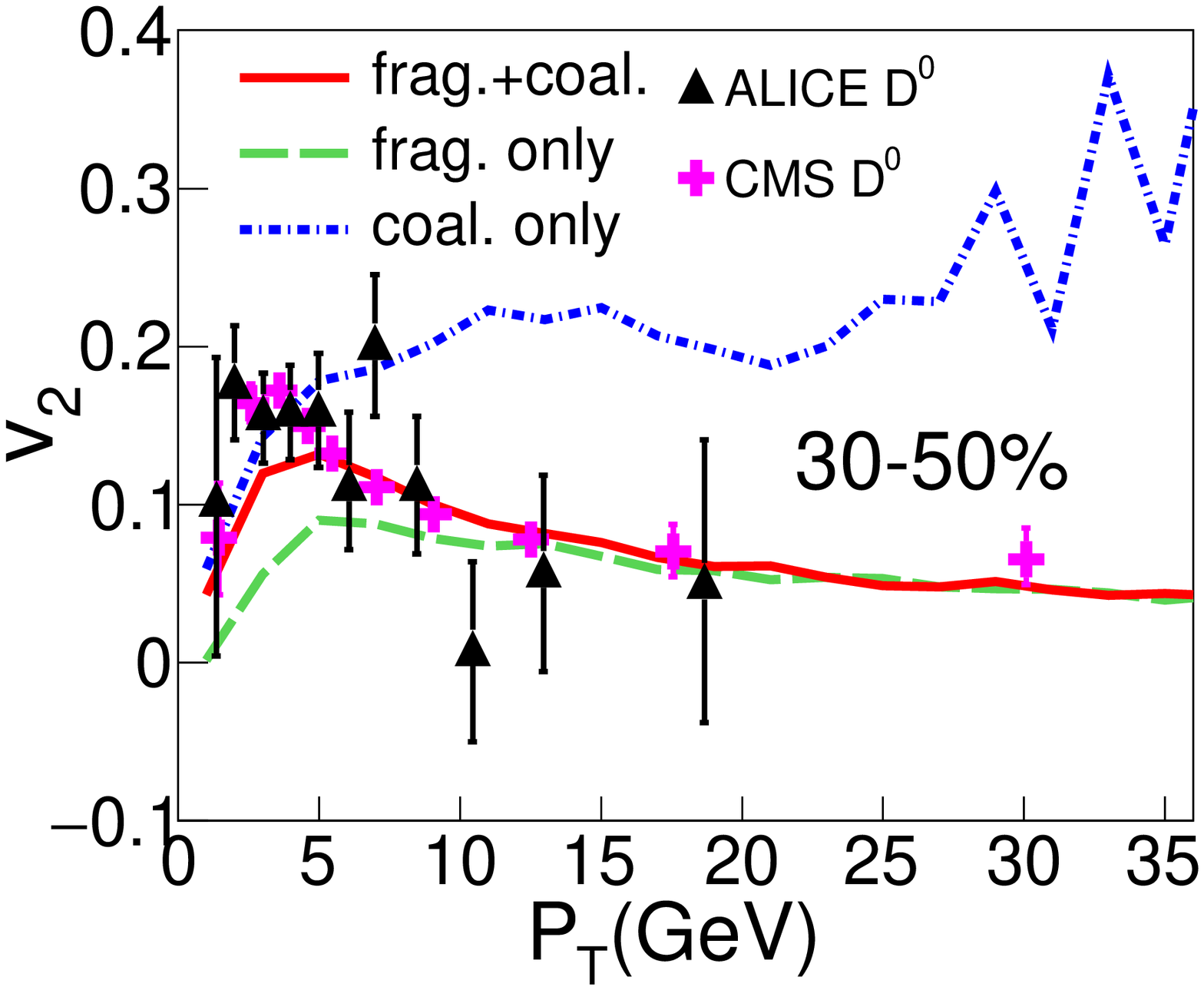}
    \caption{(Color online) Effects of different hadronization mechanisms on $D$ meson $R_\mathrm{AA}$ (left panels) and $v_2$ (right panels) at RHIC (upper panels) and the LHC (lower panels).}
    \label{fig:hadronization}
\end{figure}

In Fig.~\ref{fig:hadronization} we present contributions from different hadronization mechanisms -- fragmentation {\it vs.} coalescence -- to the heavy meson observables at RHIC and the LHC. The same parameter settings are used here as those applied in Fig.~\ref{fig:energyLoss} above. In the left panels for $R_\mathrm{AA}$, one observes that the coalescence mechanism dominates the $D$ meson yield up to $p_\mathrm{T}\sim 5$~GeV, while fragmentation dominates above that. The coalescence mechanism combines low $p_\mathrm{T}$ charm quarks with thermal light quarks into $D$ mesons, thus results in a bump structure of their $R_\mathrm{AA}$ around 2~GeV. Meanwhile, as shown in the right panels, coalescence also enhances the $D$ meson $v_2$ since it adds the larger momentum space anisotropy of thermal light quarks to charm quarks when forming $D$ mesons. At both RHIC and the LHC, introducing the coalescence mechanism is crucial for providing a reasonable description of the $D$ meson observables up to $p_\mathrm{T}\sim 10$~GeV.

\begin{figure}[tbp]
    \centering
    \includegraphics[clip=,width=0.25\textwidth]{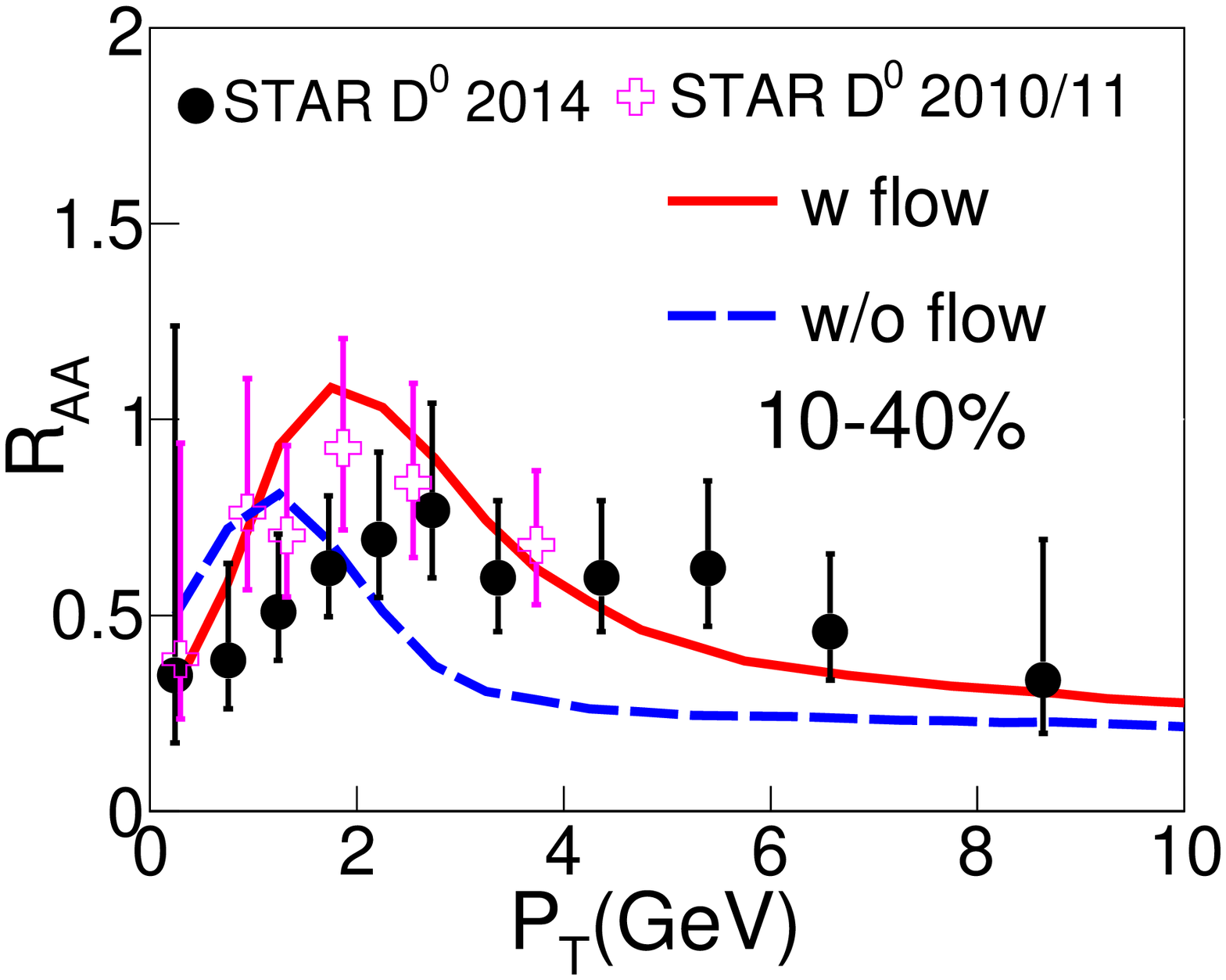}
    \hspace{-15pt}
    \includegraphics[clip=,width=0.25\textwidth]{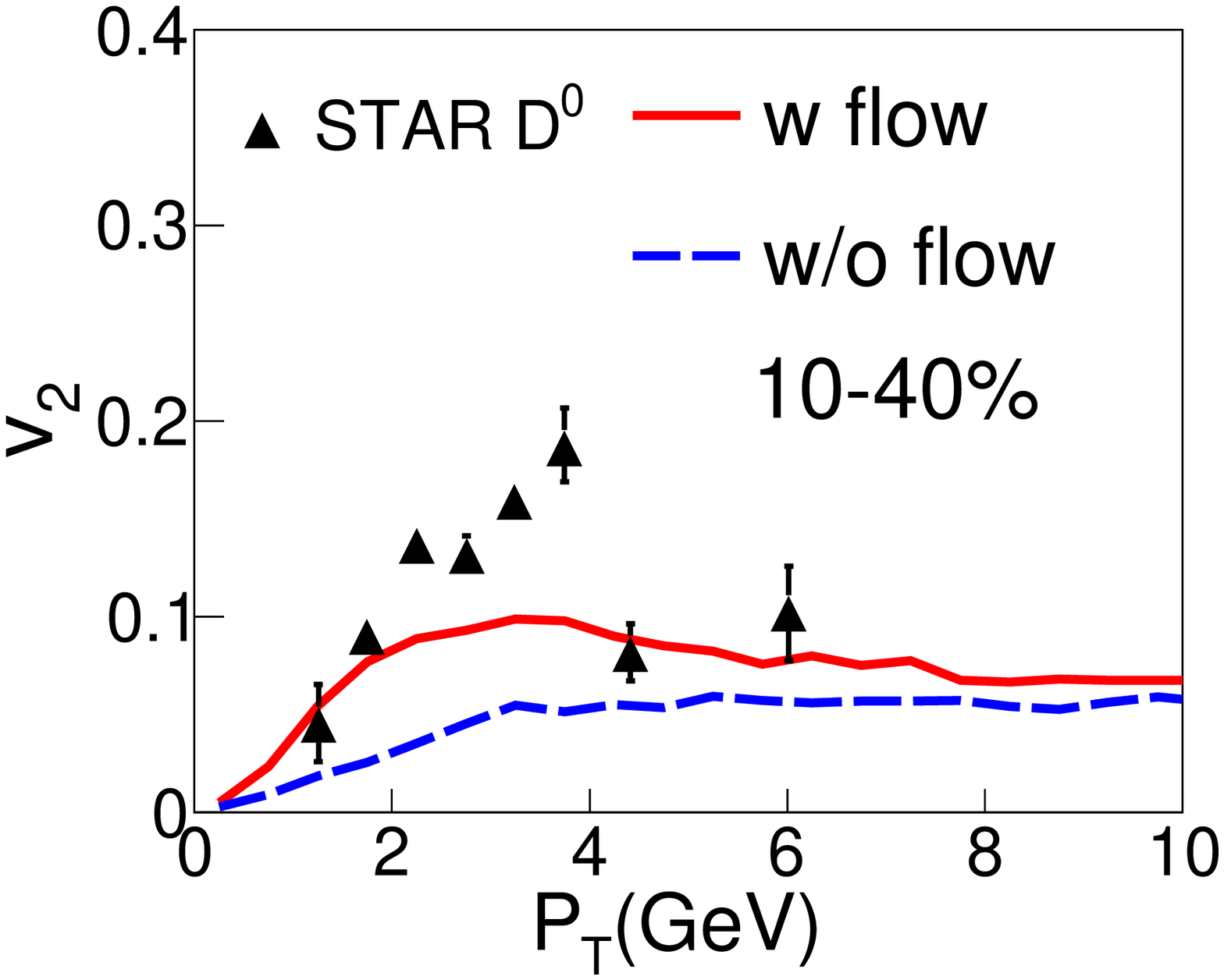}
    \includegraphics[clip=,width=0.25\textwidth]{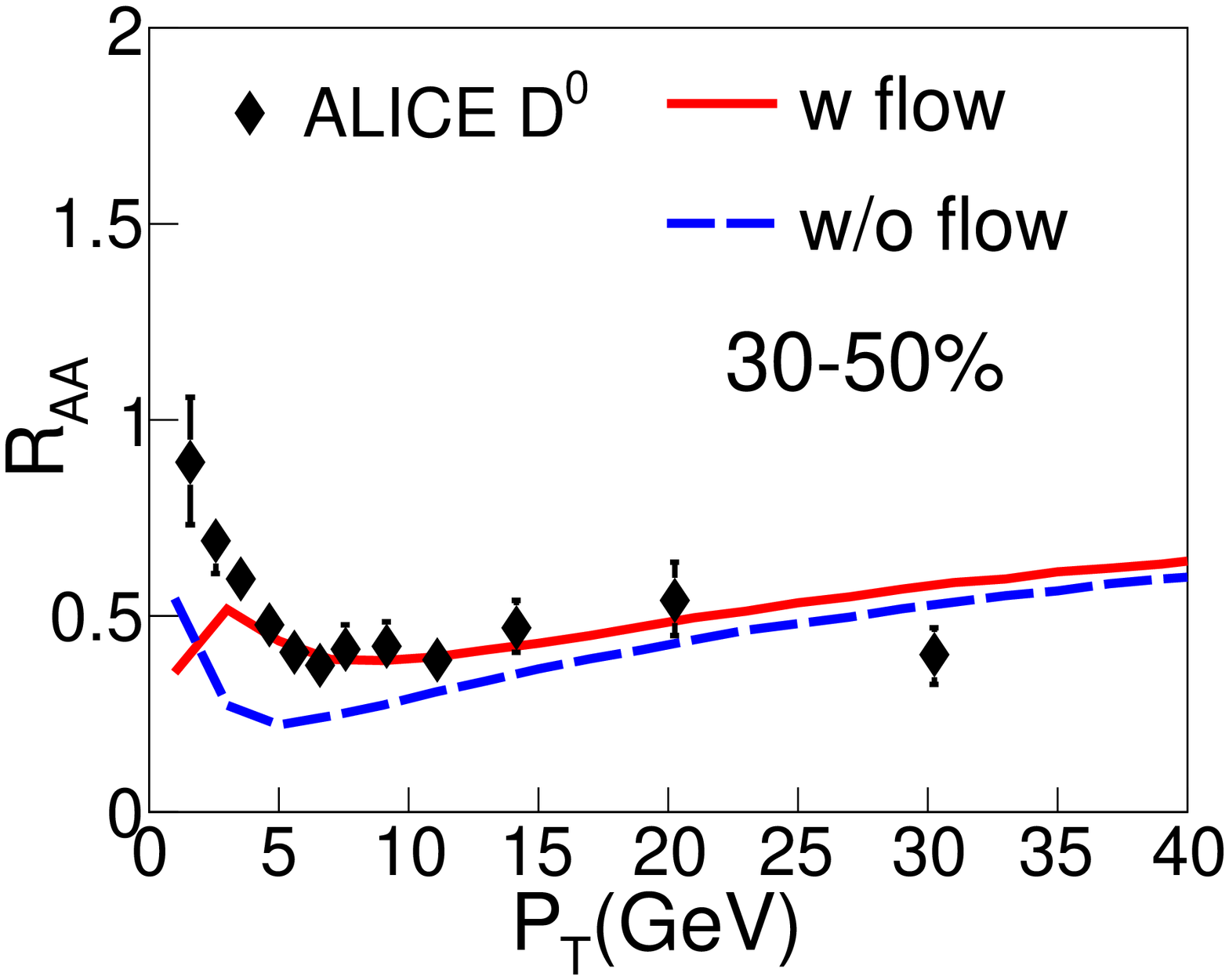}
    \hspace{-15pt}
    \includegraphics[clip=,width=0.25\textwidth]{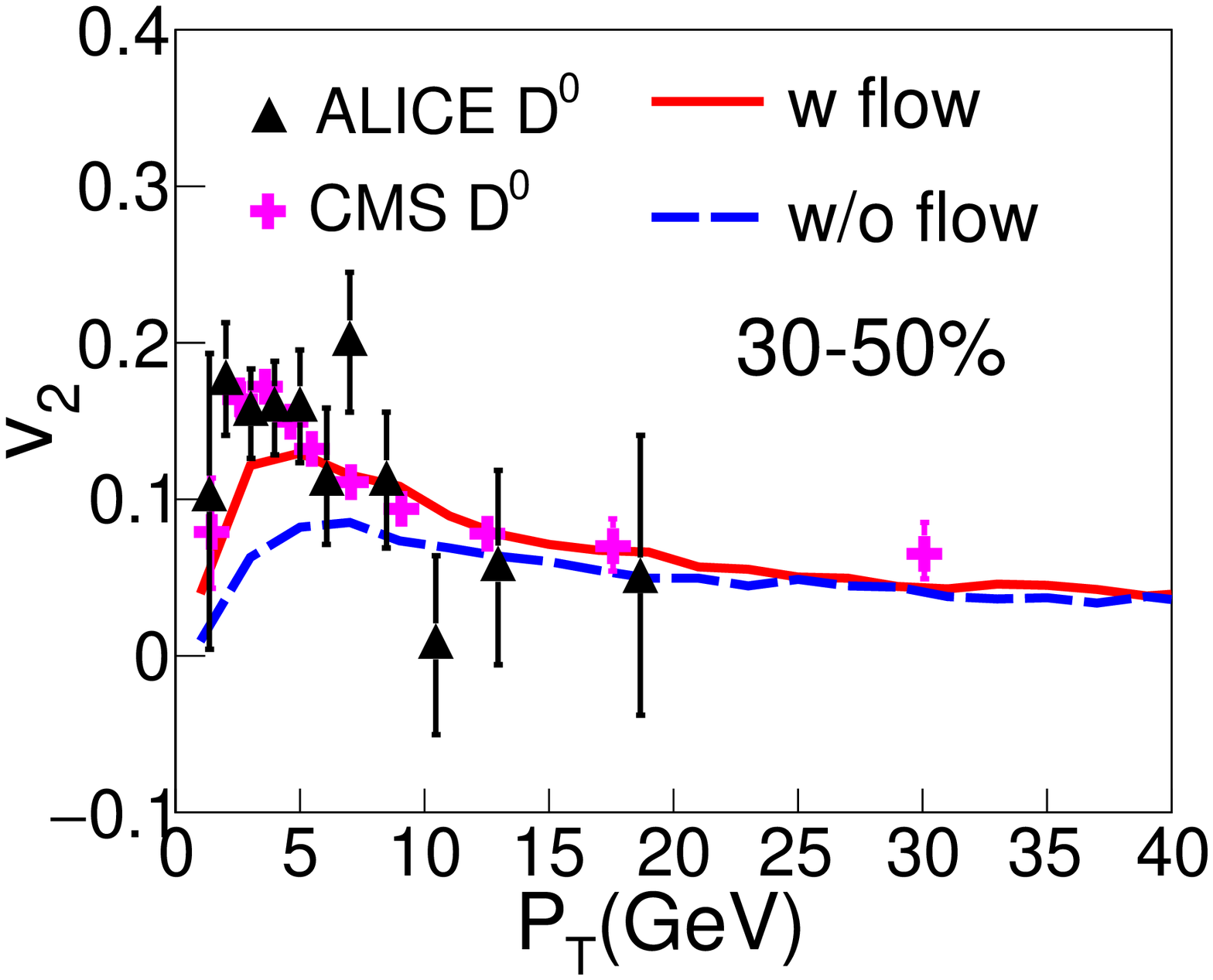}
    \caption{(Color online) Effects of QGP flow on $D$ meson $R_\mathrm{AA}$ (left panels) and $v_2$ (right panels) at RHIC (upper panels) and the LHC (lower panels).}
    \label{fig:flow}
\end{figure}

Apart from a reliable modeling of heavy quark energy loss and hadronization in QGP, the medium profile is also a crucial ingredient for describing $D$ meson observables in heavy-ion collisions. There are two aspects of medium property: geometry and radial flow. Within our Langevin-hydrodynamics model, one may switch off the impact of QGP flow on heavy quark evolution by solving the Langevin equation in the center-of-momentum frame of heavy-ion collisions instead of the local rest frame of fluid, which enable us to investigate the individual contributions from medium geometry and flow to $D$ meson observables.

In Fig.~\ref{fig:flow}, we compare the results for $D$ meson $R_\mathrm{AA}$ and $v_{2}$ with and without switching on the QGP flow. The diffusion coefficient is again set as $D_\mathrm{s}(2\pi T)=3$ (4) at RHIC (the LHC). In the left panels, one can see that the QGP flow accelerates charm quarks and thus strongly enhances $D$ meson yield (thus $R_\mathrm{AA}$) from intermediate to high $p_\mathrm{T}$ ($2\sim 10$~GeV), and meanwhile $D$ meson yield (thus $R_\mathrm{AA}$) decreases at very low $p_\mathrm{T}$. In the right panels, one can see that the effect of QGP flow on $D$ meson $v_2$ is also quite significant. At low $p_\mathrm{T}$, the anisotropic flow of QGP is the dominant source for $D$ meson $v_2$. The impact of medium flow decreases with $p_\mathrm{T}$ and becomes diminished around 20~GeV. At high $p_\mathrm{T}$, the anisotropic geometry of QGP becomes the dominant origin of $D$ meson $v_2$ since high $p_\mathrm{T}$ charm quarks lose different amount of energy when propagating along different paths through the QGP medium. The above results clearly suggest that when performing heavy and light flavor jet quenching calculations, one should use a realistic hydrodynamic simulation of the QGP evolution that is tuned to describe soft bulk observables, otherwise, the results may not be reliable.

\begin{figure}[tbp]
    \centering
    \includegraphics[clip=,width=0.25\textwidth]{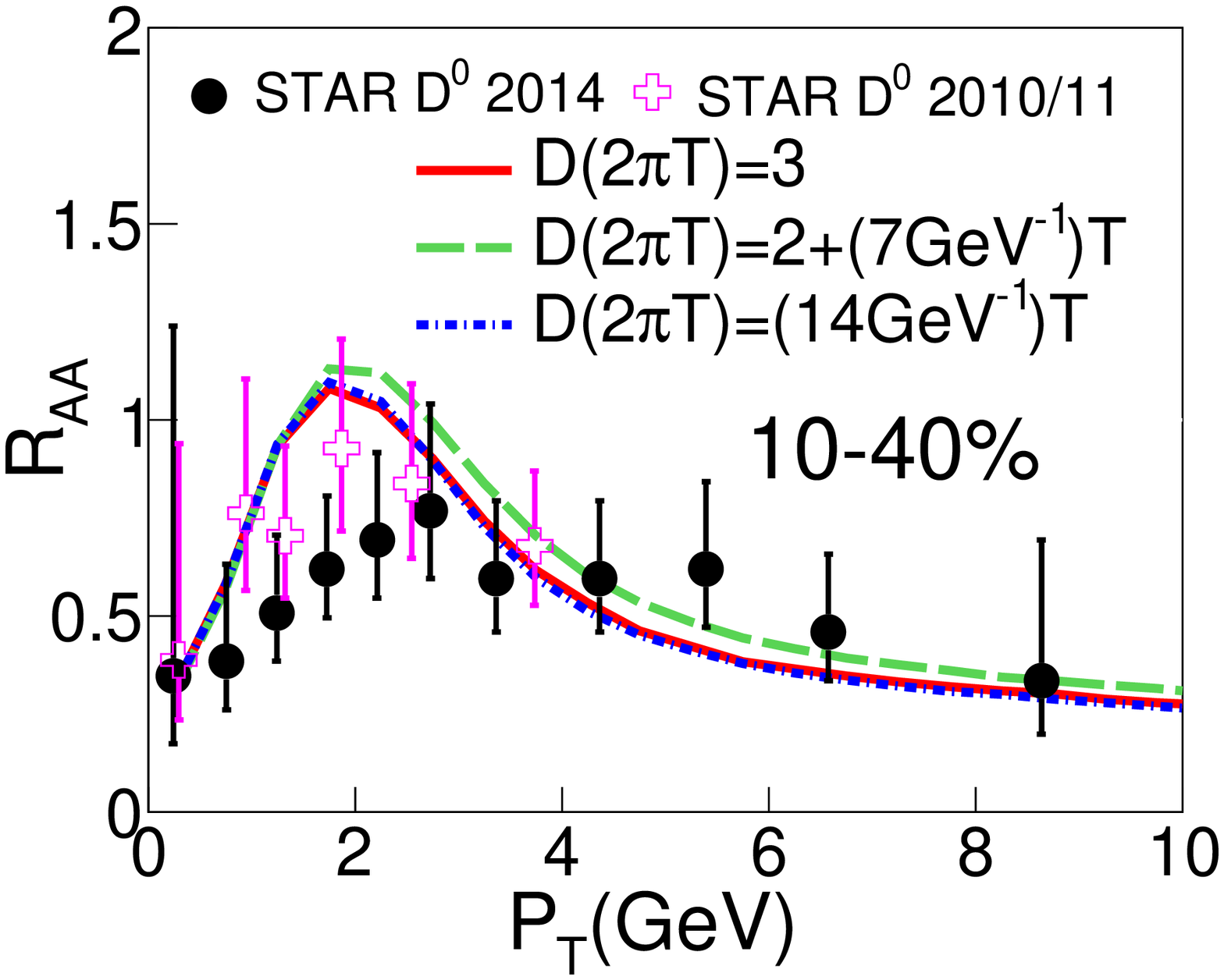}
    \hspace{-15pt}
    \includegraphics[clip=,width=0.25\textwidth]{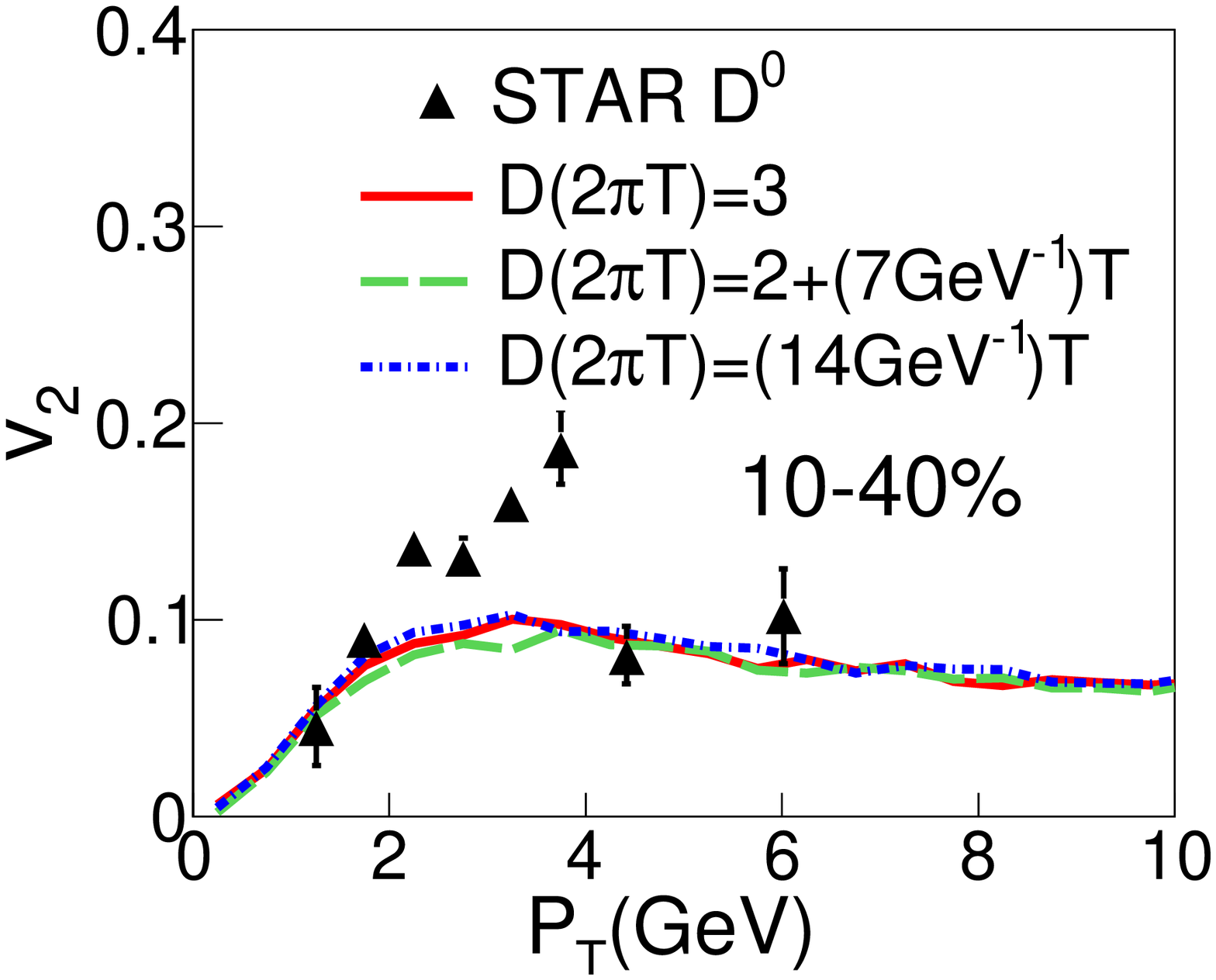}
    \includegraphics[clip=,width=0.25\textwidth]{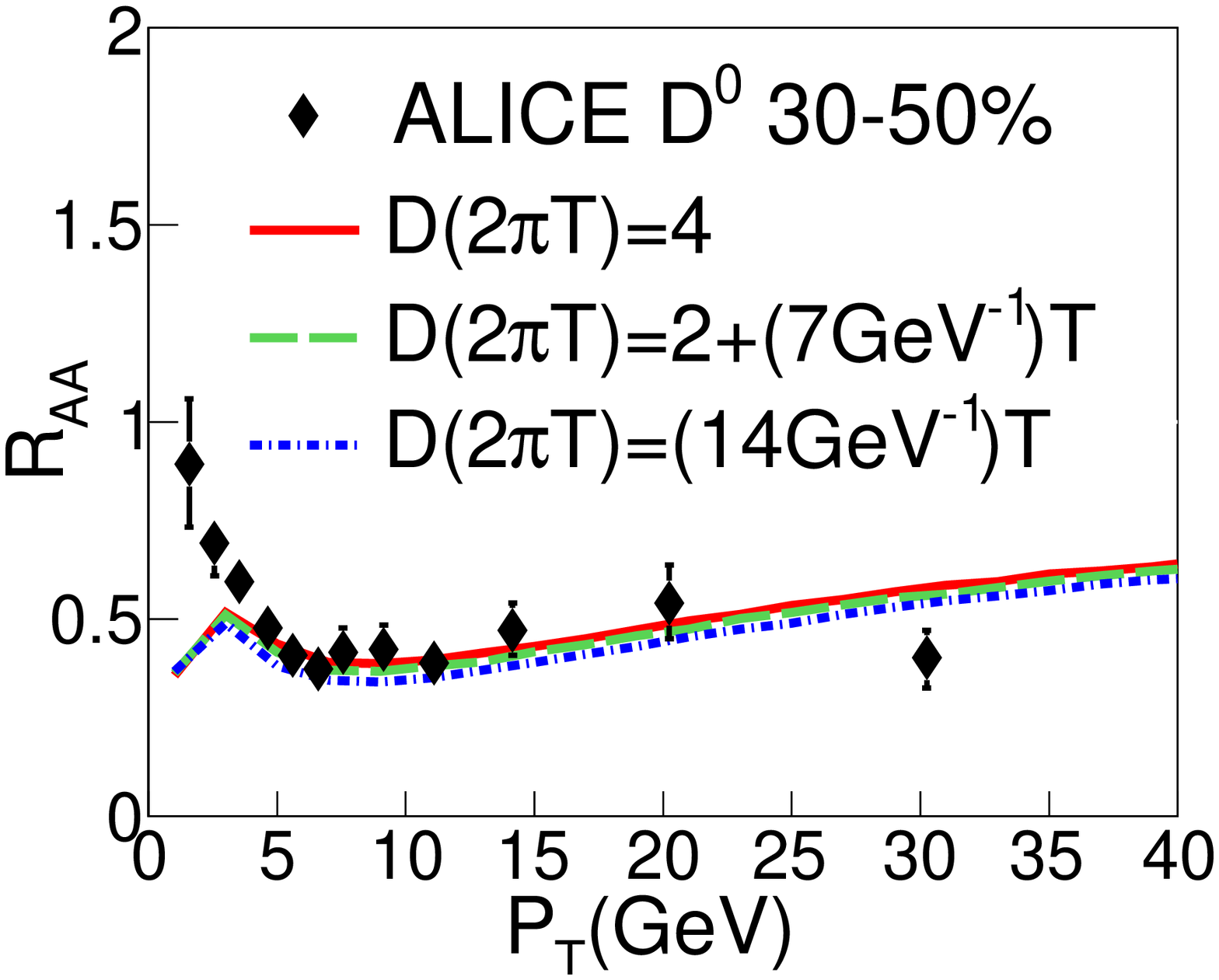}
    \hspace{-15pt}
    \includegraphics[clip=,width=0.25\textwidth]{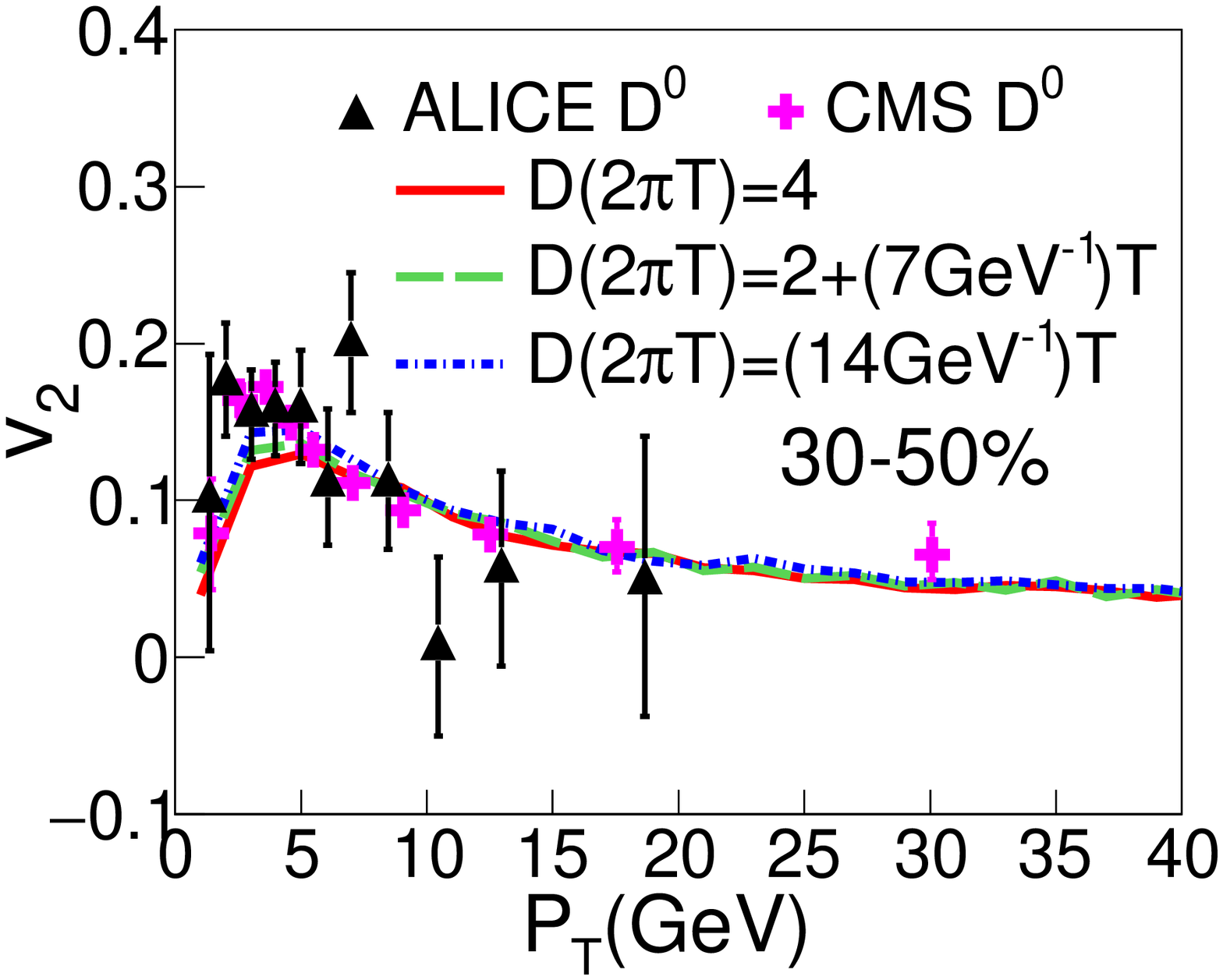}
    \caption{(Color online) Effects of the temperature dependence of the diffusion coefficient $D_\mathrm{s}(2\pi T)$ on $D$ meson $R_\mathrm{AA}$ (left panels) and $v_2$ (right panels) at RHIC (upper panels) and the LHC (lower panels).}
    \label{fig:Ds}
\end{figure}

In the end, we explore how different temperature dependences of the diffusion coefficient $D_\mathrm{s}$ affect heavy quark observables in heavy-ion collisions. Reference~\cite{Xu:2017obm} has shown that a linear dependence of $D_\mathrm{s}(2\pi T)$ on the medium temperature $T$ is a reasonable approximation, as suggested by the lattice QCD calculation and the phenomenological extraction from the model-to-data comparison. In this study, we assume the parameterized form for diffusion coefficient $D_\mathrm{s}(2\pi T)=a+bT$ and vary the slope parameter $b$ as 0, 7, and 14~GeV$^{-1}$. The parameter $a$ is adjusted such that our model provides a reasonable description of $D$ meson $R_\mathrm{AA}$ at RHIC and the LHC, as shown in the left panels of Fig.~\ref{fig:Ds}. With such setups, we compare the result using three different parameterizations of $D_\mathrm{s}(2\pi T)$: constant (3 at RHIC and 4 at the LHC), $2+(7\mathrm{GeV}^{-1})T$ and $(14\mathrm{GeV}^{-1})T$.

In the right panels of Fig.~\ref{fig:Ds}, we show the effect of different temperature dependences of $D_\mathrm{s}$ on $D$ meson $v_2$ using the above linear approximation. One can see that the effect is negligible at RHIC. At the LHC, it is within 12\% at low $p_\mathrm{T}$ and becomes invisible above around 8~GeV. It is noted that if one applies much stronger enhancement of heavy-quark-medium interaction at low temperatures (near $T_\mathrm{c}$)~\cite{Xu:2014tda,Das:2015ana}, one may observe a larger increase of elliptic flow $v_2$ for $D$ mesons as a result of more energy loss shifted to later time of the QGP evolution. However, within the linear approximation of $D_\mathrm{s}(2\pi T)$ {\it vs.} $T$, the effect on $D$ meson $v_2$ is very limited.

\section{Summary}
\label{sec:summary}

We have conducted a systematic study on heavy flavor suppression and flow in heavy-ion collisions at RHIC and the LHC. Using the state-of-the-art Langevin-hydroydnamics framework coupled to the up-to-date hybrid coalescence-fragmentation hadronization model, we have investigated in detail how various components of heavy quark model contribute to heavy meson $R_\mathrm{AA}$ and $v_2$ observed in high-energy nuclear collisions.

Our study shows that the energy loss mechanism, hadronization mechanism and medium properties are the most essential factors for correctly describing heavy flavor observables in heavy-ion collisions. While collisional energy loss dominates $D$ meson $R_\mathrm{AA}$ up to $p_\mathrm{T}= 5\sim 7$~GeV, radiative energy loss dominates at higher $R_\mathrm{AA}$. Either collisional or radiative component alone is not sufficient to provide the correct $p_\mathrm{T}$ dependence of nuclear modification of heavy flavors. The coalescence mechanism in heavy quark hadronization is crucial for describing the $D$ meson $R_\mathrm{AA}$ and $v_2$ up to $p_\mathrm{T} \sim 10$~GeV, beyond which fragmentation dominates. We have also investigated the individual contributions from two major components of the QGP -- geometry and radial flow -- on $D$ meson spectra and flow. The result clearly shows that the radial flow has a significant effect on both $R_\mathrm{AA}$ and $v_2$ of $D$ mesons below $p_\mathrm{T}\sim 10$~GeV. However, at higher $p_\mathrm{T}$, $R_\mathrm{AA}$ is mainly determined by the average temperature of medium and $v_2$ is mostly driven by the geometric anisotropy of medium.

In our work, we have also performed a systematic estimation on the uncertainties in $D$ meson suppression and flow due to different implementations of initial heavy quark spectra, heavy-quark-medium interaction in the pre-equilibrium stage and heavy quark diffusion coefficient. While different assumptions on the above aspects have little effect in the high $p_\mathrm{T}$ regime, they do introduce noticeable uncertainties at low $p_\mathrm{T}$. Applying different initial charm quark spectra (FONLL {\it vs.} LO) may yield up to 25\% difference in $R_\mathrm{AA}$ and 19\% difference in $v_2$ for $D$ mesons. The inclusion of nuclear shadowing effect can reduce $R_\mathrm{AA}$ up to 27\% at low $p_\mathrm{T}$. Delaying heavy-quark-medium interaction towards a later time can increase $D$ meson $v_2$ when the model is tuned to describe the same $R_\mathrm{AA}$. Such effect has been consistently demonstrated from three different directions in this work. (1) Delaying the starting time of heavy-quark-medium interaction from 0.6~fm to 1.2~fm can increase low $p_\mathrm{T}$ $D$ meson $v_2$ by up to 24\%. (2) The free-streaming assumption in the pre-equilibrium stage gives up to 39\% larger $v_2$ than assuming constant temperature and Bjorken evolution profiles for the pre-equilibrium stage. (3) Within the linear assumption for the temperature dependence of heavy quark diffusion coefficient $D_\mathrm{s}(2\pi T)$, one may obtain up to 12\% larger $v_2$ when increasing heavy-quark-medium interaction at low temperature. However, all these effects become negligible when $D$ meson $p_\mathrm{T}$ is above 10~GeV. These uncertainties should be considered carefully when interpreting heavy quark phenomenology or using heavy quarks to probe QGP properties in relativistic heavy-ion collisions.

\section*{Acknowledgments}

We are grateful to discussions with Weiyao Ke. This work was supported by the Natural Science Foundation of China (NSFC) under Grant Nos. 11805082, 11775095, 11890711 and 11935007, by A Project of Shandong Province Higher Educational Science of Technology Program (J17KB128), by the China Scholarship Council (CSC) under Grant No. 201906775042, by the U.S. Department of Energy (DOE) under Grant No. DE-SC0013460, and by the National Science Foundation (NSF) under Grant No. ACI-1550300 within the framework of the JETSCAPE Collaboration.

\bibliographystyle{h-physrev5}
\bibliography{SCrefs}

\end{document}